\documentclass[aps, prd, onecolumn, showpacs, showkeys, superscriptaddress, preprintnumbers, floatfix, reprint]{revtex4}
\usepackage{multirow}
\usepackage{slashed}
\usepackage{graphicx}
\usepackage{amsmath}
\usepackage{bm}
\usepackage{yhmath}
\usepackage{hyperref}
\usepackage{subfigure}
\usepackage{color}
\usepackage{cases}
\usepackage{subfigure}
\usepackage{dcolumn, booktabs, bm}
\usepackage{amsfonts, amssymb, stmaryrd, latexsym, amsmath}
\usepackage{textcomp}
\usepackage{amsthm}
\usepackage{mathrsfs}

\usepackage{array}
\newcounter{RomanNumber}

\newcommand{\lyxmathsym}[1]{\ifmmode\begingroup\def\b@ld{bold}
	\text{\ifx\math@version\b@ld\bfseries\fi#1}\endgroup\else#1\fi}

\allowdisplaybreaks
\begin{document}
	\title{Magnetic moments of the hidden-charm strange pentaquark states}
	\author{Feng  Gao}\email{feng_gao@stumail.nwu.edu.cn}\affiliation{School of Physics, Northwest University, Xian 710127, China}
	\author{Hao-Song Li}\email{haosongli@nwu.edu.cn}\affiliation{School of Physics, Northwest University, Xian 710127, China}\affiliation{Institute of Modern Physics, Northwest University, Xian 710127, China}\affiliation{Shaanxi Key Laboratory for theoretical Physics Frontiers, Xian 710127, China}\affiliation{Peng Huanwu Center for Fundamental Theory, Xian 710127, China}
	\begin{abstract}
	This paper calculates the magnetic moments of the hidden\textendash{}charm strange pentaquark states with a quantum number of  $J^P={\frac{1}{2}}^{\pm}$, ${\frac{3}{2}}^{\pm}$, ${\frac{5}{2}}^{\pm}$, and ${\frac{7}{2}}^{+}$ in the molecular, diquark\textendash{}diquark\textendash{}antiquark, and diquark\textendash{}triquark models, respectively. Numerical results demonstrate that the magnetic moments change for a different spin-orbit coupling with the same model and when involving different models with the same angular momentum.
	\end{abstract}
	\maketitle
	\thispagestyle{empty}
	\section{Introduction}
	\label{sec1}
	The quark model is a successful theory, with physicists employing it to explain mesons' and baryons' inner structures and predict the tetraquark and pentaquark. Over the past decade, the multiquark states' exploration has made significant progress both theoretically and experimentally, with several exotic hadronic states being experimentally observed ~\cite{Belle:2003nnu,LHCb:2016axx,BESIII:2016adj,LHCb:2018oeg,LHCb:2015yax,LHCb:2019kea,LHCb:2021chn,LHCb:2020jpq}.
	
	In 2015, the LHCb Collaboration observed the pentaquark states in the $J/\psi p$ invariant mass spectrum of the $\Lambda^{0}_b\to{}J/\psi{}K^{-}p{}$ decays. The two candidates of the hidden\textendash{}charm pentaquark are $P_{c}(4380)$ and $P_{c}(4450)$, respectively, whose $J^P$ has an opposite parity with $(\frac{3}{2}^{-},\frac{5}{2}^{+})$, ~\cite{LHCb:2015yax}. In 2019, the $P_{c}(4450)$ pentaquark structure was confirmed and the observations revealed comprising two peaks, $P_{c}(4440)$ and $P_{c}(4457)$, with a statistical significance of 5.4$\sigma$~\cite{LHCb:2019kea}. Meanwhile, the LHCb Collaboration reported a new pentaquark state observation, $P_{c}(4312)$, with a statistical significance of 7.3$\sigma$. In 2021, the LHCb Collaboration found evidence for a new structure $P_c(4337)$ in $B_{s}^{0}\to J/\psi p\bar{p}$ decays, with a final significance of 3.1$\sigma$~\cite{LHCb:2021chn}. The mass and width of $P_c(4337)$ are $4337\ ^{+7}_{-4}\ ^{+2}_{-2}$ MeV and $29\ ^{+26}_{-12}\ ^{+14}_{-14}$ MeV, respectively, while the parity and angular momentum of $P_{c}(4337)$ were predicted with $J^{P}\ =\ {\frac{1}{2}}^{+}$ \cite{Shen:2017ayv}.
	
	After discovering the $P_c$ states, theorists have shown great interest in explaining the pentaquark's nature. For instance, in ~\cite{Weng:2019ynv}, the authors systematically studied the mass spectrum of the $P_c$ states utilizing the chromomagnetic model, while in \cite{Wang:2016dzu}, the magnetic moments of the $P_c$ states were calculated in different color-flavor structures. In addition, the parity and angular momentum of the $P_c$ states were predicted by employing the quark delocalization color screening model ~\cite{Huang:2019jlf}. The $P_{c}(4312)$ can be identified as the hidden\textendash{}charm molecular state $\Sigma_{c}\bar{D}$ with $J^{P}=\frac{1}{2}^{-}$ and $P_{c}(4440)$, and $P_{c}(4457)$  as the hidden\textendash{}charm molecular states  $\Sigma_{c}\bar{D}^{*}$ with $J^{P}=\frac{1}{2}^{-}$ and $\frac{3}{2}^{-}$.
	
	As more exotic hadrons are observed, theorists have tried to explain their mass spectrum with the one\textendash{}boson\textendash{}exchange model~\cite{Chen:2020kco}, the QCD sum rules~\cite{Wang:2020eep, Chen:2020uif, Wang:2019got,Wang:2021itn,Ozdem:2021ugy,Azizi:2021utt}, and the effective field theory~\cite{Peng:2020hql, Lu:2021irg, Chen:2021htr, Wang:2020dko, Wang:2019nvm}. In general, pentaquark's inner structure has been classified as a molecular model~\cite{Hu:2021nvs, He:2019ify, Chen:2019asm, Wu:2019rog, Chen:2019bip, PavonValderrama:2019nbk, Fernandez-Ramirez:2019koa, Xiao:2019aya, Shimizu:2019ptd, Liu:2019tjn,Chen:2021tip,Wang:2021itn,Lu:2021irg,Xiao:2021rgp,Chen:2020uif}, diquark\textendash{}diquark\textendash{}antiquark model~\cite{Lebed:2015tna, Ali:2020vee, Wang:2019got,	 Anisovich:2017aqa, Shi:2021wyt}, and diquark\textendash{}triquark model~\cite{Karliner:2003sy, Zhu:2015bba,Giron:2021sla}.
	
	The LHCb Collaboration in 2020 observed the hidden\textendash{}charm strange pentaquark $P_{cs}(4459)$ in the $J/\psi{}\Lambda$ mass spectrum through an amplitude analysis of the $\Xi^{-}_b\to{}J/\psi{}\Lambda{}K^{-}$ decay ~\cite{LHCb:2020jpq}. The mass and width are $4458.8\pm 2.9^{+4.7}_{-1.1}$ MeV and $ 17.3\pm 6.5^{+8.0}_{-5.7}$ MeV, respectively, while after an in-depth study of $P_{cs}(4459)$, the structure was proved to have two resonances, with masses 4454.9 $\pm$ 2.7 \mbox{MeV} and 4467.8 $\pm$ 3.7 \mbox{MeV} and widths 7.5 $\pm$ 9.7 \mbox{MeV} and 5.2 $\pm$ 5.3 \mbox{MeV}, respectively. However, the parity and angular momentum of the $P_{cs}(4459)$ have not been determined experimentally. 
	The predictions of $J^{P}=\frac{1}{2}^{-}$ and $\frac{3}{2}^{-}$ have been given based on the QCD sum rules~\cite{Chen:2020uif}, chiral quark model~\cite{Hu:2021nvs},  and the strong decay behaviors of the $P_{cs}(4459)$~\cite{Chen:2021tip}.

	Given that pentaquark's magnetic moment encoding includes helpful details about the charge and magnetization distributions inside the hadrons that assists in analyzing their geometric configurations. In Ref.~\cite{Li:2021ryu}, the author studies the magnetic moments and transition magnetic moments of the hidden-charm pentaquark states with the coupled channel effects and the D wave contributions. This study is important because magnetic moments help us understand the pentaquark's inner structure. In this work, we calculate the magnetic moments of $P_{cs}$ based on the above three models. 
	
	The remainder of this paper is as follows. Sec.\ref{sec2} discusses the color factor and color configuration, while Sec.\ref{sec3} introduces the wave function of $P_{cs}$. Sec.\ref{sec4} calculates the magnetic moments of $P_{cs}$ in the molecular model,  diquark\textendash{}diquark\textendash{}antiquark model, the and diquark\textendash{}triquark model, and finally, Sec.\ref{sec5} summarizes this work.

	\section{Color factor and color configuration}
	\label{sec2}
	
The quark level involves chromomagnetic interactions. Therefore, we exploit the color factor $f$ to indicate whether the color force is attractive or repulsive. 

Regarding the quark-quark color interaction, the color factor $f$ is:
\begin{eqnarray}
	f(ik \rightarrow jl) = \frac{1}{4}\sum_{a=1}^{8}\lambda_{ji}^{a}\lambda_{lk}^{a}.  
\end{eqnarray}

where $\lambda^a$ denotes the Gell-Mann matrices and the quark colors are labelled by $i, j, k$, and $ l$. The potentials is:
\begin{eqnarray}
	V_{qq}(r) \approx +f\frac{\alpha_s}{r}. 
\end{eqnarray}
Considering the quark-antiquark color interaction, the color factor $\widetilde{f}$ is:
\begin{eqnarray}
	\widetilde{f}(ik \rightarrow jl) = -\frac{1}{4}\sum_{a=1}^{8}\lambda_{ji}^{a}\lambda_{lk}^{a}.  
\end{eqnarray}
The potentials is:
\begin{eqnarray}
	V_{q\bar{q}}(r) \approx +\widetilde{f}\frac{\alpha_s}{r}.   
\end{eqnarray}
In Table \ref{tab:magCov} we list the color factors of the multiplet in the SU(3) color representation

\begin{table}[htbp]
	\centering
	\caption{Color factor values for the color representation. }
	\label{tab:magCov} 
	\begin{tabular}{ c|c}
		\toprule[1pt]	
		\hline
		\ $3_{c}\ \otimes\ \bar{3}_{c}$\ 
		& \ \ \ $1_{c}\ \oplus\ {8}_{c}$ 
		\\
		\hline
		\ color factor\ 
		& -$\frac{4}{3}$\ \ \ \ \ \  $\frac{1}{6}$\ 
		\\
		\hline
		\ $3_{c}\ \otimes\ 3_{c}$\ 
		&  \ \ \ $6_{c}\ \oplus\ \bar{3}_{c}$ 
		\\
		\hline
		\ color factor\ 
		& \ \ \ $\frac{1}{3}$\ \ \ \ \ $-\frac{2}{3}$\ 
		\\
		\hline
		\ $3_{c}\ \otimes\ {3}_{c}\otimes\ {3}_{c}$\ 
		& \ \ \ $1_{c}\ \oplus\ {8}_{1c}\ \oplus\ {8}_{2c}\ \oplus\ {10}_{c}$ 
		\\
		\hline
		\ color factor\ 
		& $-2$\ \ \ \ \ \  $-\frac{1}{2}$\ \ \ \ \ \  $-\frac{1}{2}$\ \ \ \ \ \  $1$\ 
		\\
		\hline
		\ $3_{c}\ \otimes\ {3}_{c}\otimes\ \bar{3}_{c}$\ 
		& \ \ \ $3_{1c}\ \oplus\ 3_{2c}\ \oplus\ \bar{6}_{c}\ \oplus\ {15}_{c}$ 
		\\
		\hline
		\ color factor\ 
		& $-\frac{4}{3}$\ \ \ \ \ \  $-\frac{4}{3}$\ \ \ \ \ \  $-\frac{1}{3}$\ \ \ \ \ \  $2$\ 
		\\
		\hline
		\bottomrule[1pt]
	\end{tabular}
\end{table}	

Color confinement implies that the physical hadrons are singlet. Under this restriction, we divide the pentaquark states into the following three categories: 
\begin{enumerate}
	\item Molecular model 
	
	Each cluster of the molecular model forms a quasibound cluster. In other words, clusters of the molecular model tend to be color singlet. We observe that $f_{1_{c}} < f_{8_{c}}$ in the color representation of the quark and antiquark, hence it is easier to form a singlet state than octet states. Similarly, $f_{1_{c}} < f_{8_{1c}}/f_{8_{2c}}< f_{10_{c}}$ in the three-quark color representation and thus it is easier to form a singlet state than other states. Therefore, from the molecular model we have two configurations $(c_{}\bar{c_{}})(q_{1}q_{2}q_{3})$ and $(\bar{c_{}}q_{1})(c_{}q_{2}q_{3})$, where $q$ denotes the $u,d,s$ quark. 
	\item Diquark\textendash{}Diquark\textendash{}antiquark model 
	
	The diquark prefers to form $\bar{3}_{c}$ by comparing the color factor of $6_{c}$ and $\bar{3}_{c}$. Similarly, $\bar{3}_{c}\ \otimes\ \bar{3}_{c}$ prefers to form $3_{c}$. Hence, we have $\bar{3}_{c}(\mathcal{D})\ \otimes\ \bar{3}_{c}(\mathcal{D})\ \otimes\ \bar{3}_{c}(\mathcal{A})$ to form a color singlet, where $\mathcal{D}$ and $\mathcal{A}$ represent the diquark and antiquark, respectively. Thus, the pentaquark configuration is $(cq_{1})(q_{2}q_{3})(\bar{c})$, represented by the diquark\textendash{}diquark\textendash{}antiquark model. 
	\item Diquark\textendash{}triquark model
	
	The triquark involves two quarks and an antiquark, assisting in distinguishing it when utilizing the molecule model. In this case, $f_{3_{1c}} / f_{3_{2c}} < f_{\bar{6}_{c}} < f_{15_{c}}$ is the color representation of the triquark quark and we have $3_{c}(\mathcal{T})\ \otimes \ \bar{3}_{c}(\mathcal{D})$ to form a color singlet, where $\mathcal{T}$ represents a triquark. Thus the pentaquark configuration represented by the  diquark\textendash{}triquark model is $(c\bar{c}q_{1})(q_{2}q_{3})$ and $(cq_{1})(\bar{c}q_{2}q_{3})$ .
	
\end{enumerate}	 

The separation of c and $\bar{c}$ into distinct confinement volumes provides a natural suppression mechanism for the pentaquark widths\cite{LHCb:2019kea}. Thus we don't consider $(\bar{c}c) (q_{1}q_{2}q_{3})$ and $(\bar{c}cq_{1})(q_{2}q_{3})$. 

 	\section{Wave function of hidden-charm strange pentaquark states}
	 \label{sec3} 

		In this work, we study the pentaquark states in the $SU(3)_{f}$ frame. The overall wavefunction for a bounded multiquark state, while accounting for all degrees of freedom, can be written as:
	\begin{equation}
		\psi_{wavefunction} = \phi_{flavor}\chi_{spin}\varepsilon_{color}\eta_{space}. \nonumber
	\end{equation}
	Due to the Fermi statistics, the overall wavefunction above is required to be antisymmetric.  
	
	The molecular model of the pentaquark is made up of mesons and baryons. They have to be color singlet because of color confinement. The relation between the spin and flavor is $\phi_{flavor}\chi_{spin}$ = symmetric since the color wavefunction is antisymmetric and the spatial wavefunction is symmetric in the ground state. We study the $P_{cs}$ state in a $SU(3)_f$ frame. There are two configurations for $q_{2}q_{3}$, where $q_{2}q_{3}$ forms the $\bar{3}_f$ and $6_f$ flavor representation with the total spin S = 0 and 1, respectively. When $q_{2}q_{3}$ forms the $6_f$, it is combined with the $q_1$ to form the flavor representation  $6_f\ \otimes\ {3}_{f}$ = $10_f\ \oplus\   8_{1f}$. While, when $q_{2}q_{3}$ forms the $\bar{3}_f$, it is then combined with the $q_1$ to form the flavor representation $\bar{3}_f\ \otimes\ {3}_{f}$ = $8_{2f}\ \oplus\ 1_f$. After inserting $[c\bar{c}]$ and the Clebsch-Gordan coefficients, we apply the same method to the $(cq_{1})(q_{2}q_{3})(\bar{c})$ and $(cq_{1})(\bar{c}q_{2}q_{3})$ configurations, and we obtain the flavor wave function of $P_{cs}$ in $8_{1f}$ and $8_{2f}$. The results are reported in Table \ref{tab:wfun}.

 \begin{table}[htbp]
 	\centering
 	\caption{The flavor wave function of hidden\textendash{}charm strange pentaquark state in different models. }
 	\label{tab:wfun} 
 	\resizebox{0.79\columnwidth}{!}{
 	\begin{tabular}{c|c|c|c}
 		\toprule[1pt]	
 		\hline
 		model
 		& multiplet
 		& $(I,I_{3})$
 		& wave function
 		\\
 		\hline
 		\multirow{4}{*}{Molecular model}
 		& \multirow{2}{*}{$8_{1f}$}
 		& $(1,0)$
 		& $\frac{1}{\sqrt6}[({\bar c}d)(c\{us\})+({\bar c}u)(c\{ds\})]-\sqrt{\frac{2}{3}}({\bar c}s)(c\{ud\})$
 		\\
 		\cline{3-4} 	
 		&
 		& $(0,0)$
 		& $\frac{1}{\sqrt2}[({\bar{c}}u)(c\{ds\})-({\bar{c}}d)(c\{us\})]$
 		\\
 		\cline{2-4} 	
 		& \multirow{2}{*}{$8_{2f}$}
 		& $(1,0)$
 		& $\frac{1}{\sqrt2}\{ ({\bar c}d)(c[us])+({\bar c}u)(c[ds]) \}$
 		\\
 		\cline{3-4}
 		&
 		& $(0,0)$
 		& $\frac{1}{\sqrt6}\{({\bar c}d)(c[us])-({\bar c}u)(c[ds])-2(\bar c s)(c[ud])\}$
 		\\
 		\hline
 		\multirow{4}{*}{Diquark-diquark-antiquark model}
 		& \multirow{2}{*}{$8_{1f}$}
 		& $(1,0)$
 		& $\frac{1}{\sqrt6}[({c}d)\{us\}{\bar c}+({c}u)\{ds\}{\bar c}]-\sqrt{\frac{2}{3}}({c}s)\{ud\}{\bar c}$
 		\\
 		\cline{3-4} 	
 		&
 		& $(0,0)$
 		& $\frac{1}{\sqrt2}[({c}u)\{ds\}{\bar c}-({c}d)\{us\}{\bar c}]$
 		\\
 		\cline{2-4} 	
 		& \multirow{2}{*}{$8_{2f}$}
 		& $(1,0)$
 		& $\frac{1}{\sqrt2}\{ (cd)[us]{\bar c}+(cu)[ds]{\bar c} \}$ 
 		\\
 		\cline{3-4}
 		&
 		& $(0,0)$
 		& $\frac{1}{\sqrt6} \{ (cd)[us]{\bar c}-(cu)[ds]{\bar c}-2(cs)[ud]{\bar c} \}$
 		\\
 		\hline
 		\multirow{4}{*}{Diquark-triquark model}
 		&\multirow{2}{*}{$8_{1f}$}
 		& $(1,0)$
 		& $\frac{1}{\sqrt6}[({c}d)(\bar c\{us\})+({c}u)(\bar c\{ds\})]-\sqrt{\frac{2}{3}}({c}s)(\bar c\{ud\})$
 		\\
 		\cline{3-4} 	
 		&
 		& $(0,0)$
 		& $\frac{1}{\sqrt2}[({c}u)(\bar c\{ds\})-({c}d)(\bar c\{us\})]$
 		\\
 		\cline{2-4} 	
 		& \multirow{2}{*}{$8_{2f}$}
 		& $(1,0)$
 		& $\frac{1}{\sqrt2}\{ ({c}d)(\bar c[us])+({c}u)(\bar c[ds]) \}$
 		\\
 		\cline{3-4}
 		&
 		& $(0,0)$
 		& $\frac{1}{\sqrt6}\{({c}d)(\bar c[us])-({ c}u)(\bar c[ds])-2( c s)(\bar c[ud])\}$
 		\\
 		\hline
 		\bottomrule[1pt]
 	\end{tabular}}
 \end{table}	
	\section{Magnetic moments of hidden$\textendash{}$charm strange pentaquark}
	\label{sec4} 
	
	\subsection {Magnetic moments of the molecular model with the configuration$(\bar{c}q_1)(cq_2q_3)$}
	
	Since quarks are fundamental Dirac fermions, the operators of the total magnetic moments and the z-component are: 
	\begin{eqnarray}
		\hat{\mu}  = \ Q\frac{e}{m}\hat{S}, \ \ \ \ \ \ \    	\hat{\mu_{z}}  = \ Q\frac{e}{m}\hat{S_{z}}. 
	\end{eqnarray}
	As mentioned above, we do not consider the orbital excitation in the bound state, so the orbital excitation lies between the meson and baryon. The total magnetic moments formula can be written as:
	\begin{eqnarray}
		\hat{\mu}  = \ \hat{\mu}_{\mathcal{B}}+\hat{\mu}_{\mathcal{M}}+\hat{\mu}_{l}. 
	\end{eqnarray}
	where the subscripts $\mathcal{B}$ and $\mathcal{M}$ represent the baryon and meson, respectively, and $l$ is the orbital excitation between the meson and baryon. The magnetic moments' specific forms can be written as:
	\begin{eqnarray}
		\hat{\mu}_{\mathcal{B}}  &=& \sum_{i=1}^{3} \mu_{i}g_{i}\hat{S}_{i},\\
		\hat{\mu}_{\mathcal{M}}  &=& \sum_{i=1}^{2} \mu_{i}g_{i}\hat{S}_{i},\\
		\hat{\mu}_{l}  = \mu_{l}\hat{l} &=& \frac{M_{\mathcal{M}}\mu_{\mathcal{B}}+M_{\mathcal{B}}\mu_{\mathcal{M}}}{M_{\mathcal{M}}+M_{\mathcal{B}}}\hat{l}.
	\end{eqnarray}
	where $g_{i}$ is the Lande factor and $M_{\mathcal{M}}$ and $M_{\mathcal{B}}$ are the meson and baryon masses, respectively. The pentaquark's $(\bar{c}q_1)(cq_2q_3)$ specific magnetic moments formula in the molecular model is:
	\begin{eqnarray}
		\mu  &=& \langle\ \psi\ |\ \hat{\mu}_{\mathcal{B}}+\hat{\mu}_{\mathcal{M}}+\hat{\mu}_{l}\ |\ \psi\  \rangle\nonumber\\
		&=&
		\sum_{SS_z,ll_z}\ \langle\ SS_z,ll_z|JJ_z\ \rangle^{2}  \left \{ \mu_{l} l_z 	+
		\sum_{\widetilde{S}_\mathcal{B},\widetilde{S}_\mathcal{M}}\ \langle\ S_\mathcal{B} \widetilde{S}_{\mathcal{B}},S_\mathcal{M} \widetilde{S}_{\mathcal{M}}|SS_z\ \rangle^{2} \Bigg [
		\widetilde{S}_{\mathcal{M}}\bigg(\mu_{\bar{c}} + \mu_{q_1}\bigg )\nonumber\right.\\
		&+&\left.
		\sum_{\widetilde{S}_{c}}\ \langle\ S_c \widetilde{S}_{c},S_{r} \widetilde{S}_{\mathcal{B}}-\widetilde{S}_{c}|S_\mathcal{B} \widetilde{S}_{\mathcal{B}}\rangle^{2}\bigg(g\mu_{c}\widetilde{S}_{c}+(\widetilde{S}_{\mathcal{B}}-\widetilde{S}_{c})(\mu_{q_{2}}+\mu_{q_{3}})\bigg )
		\Bigg ]\right \}.           
	\end{eqnarray}
	where $\psi$ represents the wave function in Table\ref{tab:wfun}. $S_\mathcal{M}$, $S_\mathcal{B}$, $S_r$ are the meson, baryon, and the diquark spin inside the baryon, respectively. $\widetilde{S}$ is the third spin component.

	For example, the recently observed $P_{cs}(4459)$ state is supposed to be the $\bar{D}^{*}\Xi_{c}$ molecular states in the $8_{2f}$ representation with $(I,I_{3}) = (0,0)$. Their flavor wave functions are:
	\begin{equation}
	| P_{cs}\rangle = \frac{1}{\sqrt{6}}\{({\bar c}d)(c[us])-({\bar c}u)(c[ds])-2(\bar c s)(c[ud])\}.
	\end{equation}
	Take $J^{p}={\frac{1}{2}}^{-}$ (${\frac{1}{2}}^{+}\otimes1^{-}\otimes0^{+}$) as an example.
	$J_{1}^{P_{1}}\otimes J_{2}^{P_{2}}\otimes J_{3}^{P_{3}}$ are corresponding to the angular momentum and parity of baryon, meson and orbital, respectively. 

		\begin{align}
		\mu  & =
		\langle\ P_{cs}\ |\ \hat{\mu}_{\mathcal{B}}+\hat{\mu}_{\mathcal{M}}+\hat{\mu}_{l}\ |\  P_{cs}\  \rangle\nonumber\\
		& =
		\langle \frac{1}{2}\frac{1}{2},1 0 |\frac{1}{2}\frac{1}{2}\rangle^{2}\Bigg [
		\langle \frac{1}{2}\frac{1}{2},0 0 |\frac{1}{2}\frac{1}{2}\rangle^{2}
		\Bigg 	(\frac{1}{6}*\frac{1}{2}g\mu_{c}+\frac{1}{6}*\frac{1}{2}g\mu_{c}+\frac{4}{6}*\frac{1}{2}g\mu_{c}\Bigg )\Bigg ]
		+\nonumber\\ & 
		\ \ \ \ \ \langle \frac{1}{2}-\frac{1}{2},1 1 |\frac{1}{2}\frac{1}{2}\rangle^{2}\Bigg [	\Bigg 	(\frac{1}{6}*(\frac{1}{2}g\mu_{\bar{c}}+\frac{1}{2}g\mu_{d})+\frac{1}{6}*(\frac{1}{2}g\mu_{\bar{c}}+\frac{1}{2}g\mu_{u})+\frac{4}{6}*(\frac{1}{2}g\mu_{\bar{c}}+\frac{1}{2}g\mu_{s})\Bigg )
		+\nonumber\\ & 
		\ \ \ \ \ 	\langle \frac{1}{2}-\frac{1}{2},0 0 |\frac{1}{2}-\frac{1}{2}\rangle^{2} 	\Bigg 	(\frac{1}{6}*-\frac{1}{2}g\mu_{c}+\frac{1}{6}*-\frac{1}{2}g\mu_{c}+\frac{4}{6}*-\frac{1}{2}g\mu_{c}\Bigg )\Bigg ]
		\nonumber\\ 
		& =
		\frac{1}{9} (\mu_{u}+\mu_{d}+4\mu_{s}+6\mu_{\bar{c}}-3\mu_{c} ).
	\end{align}\\
	
	 In this work, we use the following constituent quark masses\cite{Wang:2018gpl},
	\begin{eqnarray}
		m_u \ =\ m_d \ =\  0.336\ \mbox{GeV}, \ 	m_s \ =\ 0.540\ \mbox{GeV},\  m_c \ =\ 1.660\ \mbox{GeV}. \nonumber
	\end{eqnarray}
	The numerical results with isospin $(I,I_3) = (1,0)$ and $(I,I_3) = (0,0)$ are reported in Table \ref{pcm} and \ref{lpb}, respectively.
	 
\begin{table*}[htbp]
	\caption{The magnetic moments of the pentaquark states in the molecular model with the wave function  $\frac{1}{\sqrt6}[({\bar c}d)(c\{us\})+({\bar
			c}u)(c\{ds\})]-\sqrt{\frac{2}{3}}({\bar c}s)(c\{ud\})$ in $8_{1f}$ and $\frac{1}{\sqrt2}\{ ({\bar c}d)(c[us])+({\bar c}u)(c[ds]) \}$ in $8_{2f}$ with isospin $(I,I_3) = (1,0)$. They are in $8_{1f}$ representation from $6_f \otimes 3_f = 10_f \oplus 8_{1f}$ and $8_{2f}$ representation from $\bar{3}_f \otimes 3_f = 1_f \oplus8_{2f}$, respectively. The third line $J_{1}^{P_{1}}\otimes J_{2}^{P_{2}}\otimes J_{3}^{P_{3}}$ are corresponding to the angular momentum and parity of baryon, meson and orbital, respectively. The unit is the magnetic moments of the proton.}
		\label{pcm}

	\begin{center}

		\resizebox{0.90\columnwidth}{!}{
		\begin{tabular}{c|c|c|c|c|c|c|c} \toprule[1pt]	
			\hline
			\multicolumn{8}{c}{$8_{1f}$: $\frac{1}{\sqrt6}[({\bar c}d)(c\{us\})+({\bar c}u)(c\{ds\})]-\sqrt{\frac{2}{3}}({\bar c}s)(c\{ud\})$}
			\\
			\hline
			&\multicolumn{3}{c|}{$^{2}S_{\frac{1}{2}}$	($J^P={\frac{1}{2}}^{-}$)}
			&\multicolumn{3}{c|}{${^{4}S_{\frac{3}{2}}}$(${J^P={\frac{3}{2}}^{-}}$)}
			&\multicolumn{1}{c}{$^{6}S_{\frac{5}{2}}^{-}$($J^P={\frac{5}{2}}^{-}$)} 
			\\
			\cline{2-8} 
			$(Y, I, I_3)$ 
			& ${\frac{1}{2}}^{+}\otimes0^{-}\otimes0^{+}$
			& ${\frac{1}{2}}^{+}\otimes1^{-}\otimes0^{+}$  
			& ${\frac{3}{2}}^{+}\otimes1^{-}\otimes0^{+}$      
			& ${\frac{1}{2}}^{+}\otimes1^{-}\otimes0^{+}$ 
			& ${\frac{3}{2}}^{+}\otimes0^{-}\otimes0^{+}$
			& ${\frac{3}{2}}^{+}\otimes1^{-}\otimes0^{+}$  
			& ${\frac{3}{2}}^{+}\otimes1^{-}\otimes0^{+}$
			\\
			\hline		
			$(0,1,0)$ &0.263 &-0.493 &0.735&-0.345 &0.959 &0.460 &0.352 
			\\
			\hline		
			&\multicolumn{3}{c|}{${^{2}P_{\frac{1}{2}}}$ (${J^P={\frac{1}{2}}^{+}}$)} &\multicolumn{3}{c|}{${^{4}P_{\frac{1}{2}}}$ (${J^P={\frac{1}{2}}^{+}}$)}  
			&
			\\
			\cline{2-8}
			$(Y, I, I_3)$ 
			& ${\frac{1}{2}}^{+}\otimes0^{-}\otimes1^{-}$  
			& $[{\frac{1}{2}}^{+}\otimes1^{-}]_{\frac{1}{2}}\otimes1^{-}$      
			& $[{\frac{3}{2}}^{+}\otimes1^{-}]_{\frac{1}{2}}\otimes1^{-}$  
			& ${\frac{3}{2}}^{+}\otimes0^{-}\otimes1^{-}$    
			& $[{\frac{1}{2}}^{+}\otimes1^{-}]_{\frac{3}{2}}\otimes1^{-}$  
			& $[{\frac{3}{2}}^{+}\otimes1^{-}]_{\frac{3}{2}}\otimes1^{-}$
			
			\\
			\hline
			$(0,1,0)$ &-0.145&0.125 &-0.289&0.564 &-0.172 &0.278&  
			\\			
			\hline			
			&\multicolumn{3}{c|}{${^{2}P_{\frac{3}{2}}}$(${J^P={\frac{3}{2}}^{+}}$)}
			&\multicolumn{3}{c|}{${^{4}P_{\frac{3}{2}}}$(${J^P={\frac{3}{2}}^{+}}$)}
			&\multicolumn{1}{c}{${^{6}P_{\frac{3}{2}}}$(${J^P={\frac{3}{2}}^{+}}$)}  
			\\
			\cline{2-8}
			$(Y, I, I_3)$ 
			& ${\frac{1}{2}}^{+}\otimes0^{-}\otimes1^{-}$  
			&  $[{\frac{1}{2}}^{+}\otimes1^{-}]_{\frac{1}{2}}\otimes1^{-}$      
			& $[{\frac{3}{2}}^{+}\otimes1^{-}]_{\frac{1}{2}}\otimes1^{-}$   &${\frac{3}{2}}^{+}\otimes0^{-}\otimes1^{-}$    
			& $[{\frac{1}{2}}^{+}\otimes1^{-}]_{\frac{3}{2}}\otimes1^{-}$
			& $[{\frac{3}{2}}^{+}\otimes1^{-}]_{\frac{3}{2}}\otimes1^{-}$   
			& $[{\frac{3}{2}}^{+}\otimes1^{-}]_{\frac{5}{2}}\otimes1^{-}$
			\\
			\hline			
			$(0,1,0)$ &0.177 &-0.551 &0.669&0.666 &-0.276 &0.311 &0.335 
			\\			
			\hline			
			&\multicolumn{3}{c|}{${^{4}P_{\frac{5}{2}}}$ (${J^P={\frac{5}{2}}^{+}}$)} &\multicolumn{1}{c|}{${^{6}P_{\frac{5}{2}}}$ (${J^P={\frac{5}{2}}^{+}}$)}  &\multicolumn{1}{c|}{${^{6}P_{\frac{7}{2}}}$ (${J^P={\frac{7}{2}}^{+}}$)}
			\\
			\cline{2-8}
			$ (Y, I, I_3)$  
			& ${\frac{1}{2}}^{+}\otimes1^{-}\otimes1^{-}$ 
			& ${\frac{3}{2}}^{+}\otimes0^{-}\otimes1^{-}$    
			& $[{\frac{3}{2}}^{+}\otimes1^{-}]_{\frac{3}{2}}\otimes1^{-}$  
			& $[{\frac{3}{2}}^{+}\otimes1^{-}]_{\frac{5}{2}}\otimes1^{-}$  
			& ${\frac{3}{2}}^{+}\otimes1^{-}\otimes1^{-}$
			\\
			\hline			
			$(0,1,0)$  &-0.403 &0.865 &0.394 &0.292 &0.285 
			\\
			\hline
			\bottomrule[1pt] 
			\multicolumn{8}{c}{$8_{2f}$: $\frac{1}{\sqrt2}\{ ({\bar c}d)(c[us])+({\bar c}u)(c[ds]) \}$}
			\\
			\hline
			&\multicolumn{2}{c|}{$^2S_{\frac{1}{2}}$($J^P={\frac{1}{2}}^{-}$)}
			&\multicolumn{1}{c|}{$^{4}S{{\frac{3}{2}}}$(${J^P={\frac{3}{2}}^{-}}$)}
			&\multicolumn{2}{c|}{${^{2}P_{\frac{1}{2}}}$(${J^P={\frac{1}{2}}^{+}}$)}
			&\multicolumn{1}{c|}{${^{4}P_{\frac{1}{2}}}$(${J^P={\frac{1}{2}}^{+}}$)}
			\\
			\cline{2-8} 
			$(Y, I, I_3)$  
			& ${\frac{1}{2}}^{+}\otimes0^{-}\otimes0^{+}$
			& ${\frac{1}{2}}^{+}\otimes1^{-}\otimes0^{+}$
			& ${\frac{1}{2}}^{+}\otimes1^{-}\otimes0^{+}$
			& ${\frac{1}{2}}^{+}\otimes0^{-}\otimes1^{-}$
			& $[{\frac{1}{2}}^{+}\otimes1^{-}]_{\frac{1}{2}}\otimes1^{-}$ 
			& $[{\frac{1}{2}}^{+}\otimes1^{-}]_{\frac{3}{2}}\otimes1^{-}$ 
			\\
			\hline			
			$(0,1,0)$ &0.377&-0.067&0.465&-0.167&-0.007&0.273
			\\			
			\midrule[1pt]									
			&\multicolumn{2}{c|}{$^{2}P_{\frac{3}{2}}$ ($J^P={\frac{3}{2}}^{+}$)} &\multicolumn{1}{c|}{$^{4}P_{\frac{3}{2}}$ ($J^P={\frac{3}{2}}^{+}$)}   &\multicolumn{1}{c|}{${^{4}P_{\frac{5}{2}}^{+}}$ (${J^P={\frac{5}{2}}^{+}}$)}   
			\\
			\cline{2-8}
			$(Y, I, I_3)$ 
			& ${\frac{1}{2}}^{+}\otimes0^{-}\otimes1^{-}$
			& $[{\frac{1}{2}}^{+}\otimes0^{-}]_{\frac{3}{2}}\otimes1^{-}$ 
			& ${\frac{1}{2}}^{+}\otimes1^{-}\otimes1^{-}$
			& ${\frac{1}{2}}^{+}\otimes1^{-}\otimes1^{-}$ 
			\\
			\hline			
			$(0,1,0)$ &0.315&-0.110&0.324&0.422\\					
			\bottomrule[1pt]
		\end{tabular}}
	\end{center}
\end{table*}
\begin{table*}[htbp]
	\caption{The magnetic moments of the pentaquark states in the molecular model with the wave function  $\frac{1}{\sqrt2}[({\bar{c}}u)(c\{ds\})-({\bar{c}}d)(c\{us\})]$ in $8_{1f}$ and $\frac{1}{\sqrt6} \{ ({\bar c}d)(c[us])-({\bar c}u)(c[ds])-2(\bar c s)(c[ud]) \}$ in $8_{2f}$ with isospin $(I,I_3) = (0,0)$. The third line $J_{1}^{P_{1}}\otimes J_{2}^{P_{2}}\otimes J_{3}^{P_{3}}$ are corresponding to the angular momentum and parity of baryon, meson and orbital, respectively. The unit is the magnetic moments of the proton.}\label{lpb}
	\begin{center}
		\resizebox{0.90\columnwidth}{!}{
			\begin{tabular}{c|c|c|c|c|c|c|c} \toprule[1pt]	
				\hline
				\multicolumn{8}{c}{$8_{1f}$: $\frac{1}{\sqrt2}[({\bar{c}}u)(c\{ds\})-({\bar{c}}d)(c\{us\})]$}
				\\
				\hline
				&\multicolumn{3}{c|}{$^{2}S_{\frac{1}{2}}$($J^P={\frac{1}{2}}^{-}$)}
				&\multicolumn{3}{c|}{${^{4}S_{\frac{3}{2}}}$(${J^P={\frac{3}{2}}^{-}}$)}
				&\multicolumn{1}{c}{$^{6}S_{\frac{5}{2}}^{-}$($J^P={\frac{5}{2}}^{-}$)} 
				\\
				\cline{2-8} 
				$(Y, I, I_3)$ 
				& ${\frac{1}{2}}^{+}\otimes0^{-}\otimes0^{+}$
				& ${\frac{1}{2}}^{+}\otimes1^{-}\otimes0^{+}$ 
				& ${\frac{3}{2}}^{+}\otimes1^{-}\otimes0^{+}$      
				& ${\frac{1}{2}}^{+}\otimes1^{-}\otimes0^{+}$
				& ${\frac{3}{2}}^{+}\otimes0^{-}\otimes0^{+}$
				& ${\frac{3}{2}}^{+}\otimes1^{-}\otimes0^{+}$  
				& ${\frac{3}{2}}^{+}\otimes1^{-}\otimes0^{+}$
				\\
				\hline				
				$(0,0,0)$ &-0.201&0.126 &0.117&-0.113 &0.263 &0.228 &0.352 
				\\
				\hline			
				&\multicolumn{3}{c|}{${^{2}P_{\frac{1}{2}}}$ (${J^P={\frac{1}{2}}^{+}}$)} &\multicolumn{3}{c|}{${^{4}P_{\frac{1}{2}}}$ (${J^P={\frac{1}{2}}^{+}}$)}
				&  
				\\
				\cline{2-8}
				$(Y, I, I_3)$ 
				& ${\frac{1}{2}}^{+}\otimes0^{-}\otimes1^{-}$  
				& $[{\frac{1}{2}}^{+}\otimes1^{-}]_{\frac{1}{2}}\otimes1^{-}$      
				& $[{\frac{3}{2}}^{+}\otimes1^{-}]_{\frac{1}{2}}\otimes1^{-}$  
				& ${\frac{3}{2}}^{+}\otimes0^{-}\otimes1^{-}$    
				& $[{\frac{1}{2}}^{+}\otimes1^{-}]_{\frac{3}{2}}\otimes1^{-}$  
				& $[{\frac{3}{2}}^{+}\otimes1^{-}]_{\frac{3}{2}}\otimes1^{-}$
				&
				\\
				\hline
				$(0,0,0)$ &0.021 &-0.076 &-0.076&-0.046 &0.171 &0.145 & 
				\\				
				\hline			
				&\multicolumn{3}{c|}{${^{2}P_{\frac{3}{2}}}$(${J^P={\frac{3}{2}}^{+}}$)}
				&\multicolumn{3}{c|}{${^{4}P_{\frac{3}{2}}}$(${J^P={\frac{3}{2}}^{+}}$)}
				&\multicolumn{1}{c}{${^{6}P_{\frac{3}{2}}}$	(${J^P={\frac{3}{2}}^{+}}$)}  \\
				\cline{2-8}
				$(Y, I, I_3)$ 
				& ${\frac{1}{2}}^{+}\otimes0^{-}\otimes1^{-}$ 
			    & $[{\frac{1}{2}}^{+}\otimes1^{-}]_{\frac{1}{2}}\otimes1^{-}$     
				& $[{\frac{3}{2}}^{+}\otimes1^{-}]_{\frac{1}{2}}\otimes1^{-}$   
				& ${\frac{3}{2}}^{+}\otimes0^{-}\otimes1^{-}$    
				& $[{\frac{1}{2}}^{+}\otimes1^{-}]_{\frac{3}{2}}\otimes1^{-}$ 
				& $[{\frac{3}{2}}^{+}\otimes1^{-}]_{\frac{3}{2}}\otimes1^{-}$  
				& $[{\frac{3}{2}}^{+}\otimes1^{-}]_{\frac{5}{2}}\otimes1^{-}$
				\\
				\hline				
				$(0,0,0)$ &-0.270 &0.075 &0.061&-0.103 &0.163 &0.145 &0.329 
				\\			
				\hline
				&\multicolumn{3}{c|}{${^{4}P_{\frac{5}{2}}}$ (${J^P={\frac{5}{2}}^{+}}$)} &\multicolumn{1}{c|}{${^{6}P_{\frac{5}{2}}}$ (${J^P={\frac{5}{2}}^{+}}$)}  &\multicolumn{1}{c|}{${^{6}P_{\frac{7}{2}}}$ (${J^P={\frac{7}{2}}^{+}}$)}
				\\
				\cline{2-8}
				$ (Y, I, I_3)$   
				& ${\frac{1}{2}}^{+}\otimes1^{-}\otimes1^{-}$ 
				& ${\frac{3}{2}}^{+}\otimes0^{-}\otimes1^{-}$    
				& $[{\frac{3}{2}}^{+}\otimes1^{-}]_{\frac{3}{2}}\otimes1^{-}$   
				& $[{\frac{3}{2}}^{+}\otimes1^{-}]_{\frac{5}{2}}\otimes1^{-}$   
				& ${\frac{3}{2}}^{+}\otimes1^{-}\otimes1^{-}$
				\\
				\hline
				
				$(0,0,0)$  &-0.164 &0.189 &0.172 &0.295 &0.296 
				\\
				\hline
				\bottomrule[1pt]
				\multicolumn{8}{c}{$8_{2f}$: $\frac{1}{\sqrt6} \{ ({\bar c}d)(c[us])-({\bar c}u)(c[ds])-2(\bar c s)(c[ud]) \}$}
				\\
				\hline
				&\multicolumn{2}{c|}{$^2S_{\frac{1}{2}}$($J^P={\frac{1}{2}}^{-}$)}
				&\multicolumn{1}{c|}{$^{4}S{{\frac{3}{2}}}$(${J^P={\frac{3}{2}}^{-}}$)}
				&\multicolumn{2}{c|}{${^{2}P_{\frac{1}{2}}}$(${J^P={\frac{1}{2}}^{+}}$)}
				&\multicolumn{1}{c|}{${^{4}P_{\frac{1}{2}}}$(${J^P={\frac{1}{2}}^{+}}$)}
				\\
				\cline{2-8} 
				$(Y, I, I_3)$  
				& ${\frac{1}{2}}^{+}\otimes0^{-}\otimes0^{+}$
				& ${\frac{1}{2}}^{+}\otimes1^{-}\otimes0^{+}$
				& ${\frac{1}{2}}^{+}\otimes1^{-}\otimes0^{+}$
				& ${\frac{1}{2}}^{+}\otimes0^{-}\otimes1^{-}$
				& $[{\frac{1}{2}}^{+}\otimes1^{-}]_{\frac{1}{2}}\otimes1^{-}$ 
				& $[{\frac{1}{2}}^{+}\otimes1^{-}]_{\frac{3}{2}}\otimes1^{-}$ 
				\\
				\hline			
				$(0,0,0)$ &0.377&-0.531&-0.231&-0.161&0.152&-0.116
				\\			
				\hline									
				&\multicolumn{2}{c|}{$^{2}P_{\frac{3}{2}}$ ($J^P={\frac{3}{2}}^{+}$)} &\multicolumn{1}{c|}{$^{4}P_{\frac{3}{2}}$ ($J^P={\frac{3}{2}}^{+}$)}   &\multicolumn{1}{c|}{${^{4}P_{\frac{5}{2}}^{+}}$(${J^P={\frac{5}{2}}^{+}}$}\\
				\cline{2-8}
				$(Y, I, I_3)$ 
				& ${\frac{1}{2}}^{+}\otimes0^{-}\otimes1^{-}$
				& $[{\frac{1}{2}}^{+}\otimes0^{-}]_{\frac{3}{2}}\otimes1^{-}$ 
				& ${\frac{1}{2}}^{+}\otimes1^{-}\otimes1^{-}$
				& ${\frac{1}{2}}^{+}\otimes1^{-}\otimes1^{-}$ 
				\\
				\hline			
				$(0,0,0)$ &0.324&-0.568&-0.184&-0.268
				\\	
				\hline				
				\bottomrule[1pt]
		\end{tabular}}
	\end{center}
\end{table*}

	\subsection {Magnetic moments of the diquark-diquark-antiquark model with the $(cq_1)(q_2q_3)\bar{c}$ configuration}

In the diquark-diquark-antiquark model, there are two P-wave excitation modes inside the three-body bound state, the $\rho$ and the $\lambda$ excitation. The $\rho$ mode P-wave orbital excitation lies between the diquark $(cq_1)$ and diquark $(q_2q_3)$. The $\lambda$ mode P-wave orbital excitation lies between the $\bar{c}$ and the center of mass system of the $(cq_1)$ and $(q_2q_3)$.

The total magnetic moments formula of the diquark-diquark-antiquark model can be written as:
\begin{eqnarray}
	\hat{\mu}  = \ \hat{\mu}_{H}+\hat{\mu}_{L}+\hat{\mu}_{\bar{c}}+\hat{\mu}_{l}. 
\end{eqnarray}
where the subscripts $H$ and $L$ represent a heavy diquark $(cq_1)$ and light diquark $(q_2q_3)$, respectively, and $l$ is the orbital excitation. In the diquark-diquark-antiquark model, the specific magnetic moments formula of the pentaquark $(cq_1)(q_2q_3)\bar{c}$ is: 
\begin{eqnarray}
	\mu  &=& \langle\ \psi \ |\ \hat{\mu}_{H}+\hat{\mu}_{L}+\hat{\mu}_{\bar{c}}+\hat{\mu}_{l}\ |\ \psi \  \rangle\nonumber\\
	&=&\sum_{S_z,l_z}\ \langle\ SS_z,ll_z|JJ_z\ \rangle^{2}  \left \{ \mu_{l} l_z + \sum_{\widetilde{S}_{\bar{c}}}\ \langle\ S_{\bar{c}} \widetilde{S}_{\bar{c}},S_{\mathcal{G}} \widetilde{S}_{\mathcal{G}}|SS_z\ \rangle^{2} 	\Bigg [g\widetilde{S}_{\bar{c}}\mu_{\bar{c}}\nonumber\right.\\
	&+&\left.\sum_{\widetilde{S}_{H},\widetilde{S}_{L}}\ \langle\ S_{H} \widetilde{S}_{H},S_{L} \widetilde{S}_{L}|S_{\mathcal{G}} \widetilde{S}_{\mathcal{G}}\rangle^{2}\bigg(\widetilde{S}_{H}(\mu_{c}+\mu_{q_1})+\widetilde{S}_{L}(\mu_{q_2}+\mu_{q_3})\bigg )	\Bigg ]\right \}.        
\end{eqnarray}
where $S_{\mathcal{G}}$ represents the spin of $(cq_1)(q_2q_3)$. The diquarks' masses are \cite{Ebert:2010af}:
\begin{eqnarray}
	[u,d]&=& 710\mbox{MeV}, \ \ \ \ \{u,d\} =909\mbox{MeV},\ \ \ \ \ [u,s]=948\mbox{MeV},\ \ \ \  \{u,s\} =1069\mbox{MeV},\nonumber\\ \nonumber
	[c,q]&=& 1973\mbox{MeV},\ \ \ \{c,q\} =2036\mbox{MeV},\ \ \ \ [c,s]=2091\mbox{MeV},\ \ \ \{c,s\} =2158\mbox{MeV}.
\end{eqnarray}
The numerical results for the states with the $\rho$ excitation mode  with isospin $(I,I_3) = (1,0)$ and $(I,I_3) = (0,0)$ are presented in Table \ref{caqs} and \ref{qag}, respectively. 
The numerical results for the states with the $\lambda$ excitation mode  with isospin $(I,I_3) = (1,0)$ and $(I,I_3) = (0,0)$ are presented in Table \ref{abc} and \ref{def}, respectively. 
	
	\begin{table*}[htbp]
	\caption{The magnetic moments of the pentaquark states in the diquark-diquark-antiquark model with the wave function  $\frac{1}{\sqrt6}[({c}d)\{us\}{\bar c}+({c}u)\{ds\}{\bar c}]-\sqrt{\frac{2}{3}}({c}s)\{ud\}{\bar c}$ in $8_{1f}$ and $\frac{1}{\sqrt2}\{ (cd)[us]{\bar c}+(cu)[ds]{\bar c} \}$ in $8_{2f}$ with isospin $(I,I_3) = (1,0)$. They are in $8_{1f}$ representation from $6_f \otimes 3_f = 10_f \oplus 8_{1f}$ and $8_{2f}$ representation from $\bar{3}_f \otimes 3_f = 1_f \oplus8_{2f}$, respectively. The third line $J_{1}^{P_{1}}\otimes J_{2}^{P_{2}}\otimes J_{3}^{P_{3}}\otimes J_{4}^{P_{4}}$ are corresponding to the angular momentum and parity of $(cq_1)$, $(q_2q_3)$, $\bar{c}$ and orbital, respectively.The $\rho$ mode P-wave orbital excitation lies between the diquark $(cq_1)$ and diquark $(q_2q_3)$. The unit is the magnetic moments of the proton.}
	\label{caqs}
	\begin{center}
		\resizebox{0.95\columnwidth}{!}{
			\begin{tabular}{c|c|c|c|c|c|c|c} \toprule[1pt]
				\hline
				\multicolumn{8}{c} {$8_{1f}$: $\frac{1}{\sqrt6}[({c}d)\{us\}{\bar c}+({c}u)\{ds\}{\bar c}]-\sqrt{\frac{2}{3}}({c}s)\{ud\}{\bar c}$}
				\\
				\hline
				&\multicolumn{3}{c|}{$^{2}S_{\frac{1}{2}}$($J^P={\frac{1}{2}}^{-}$)}
				&\multicolumn{3}{c|}{${^{4}S_{\frac{3}{2}}}$(${J^P={\frac{3}{2}}^{-}}$)}
				&\multicolumn{1}{c}{$^{6}S_{\frac{5}{2}}$ ($J^P={\frac{5}{2}}^{-}$)}
				\\
				\cline{2-8}
				$(Y, I, I_3)$   
				& $0^{+}\otimes1^{+} \otimes{\frac{1}{2}}^{-}\otimes0^{+}$
				&$(1^{+}\otimes1^{+})_{0} \otimes {\frac{1}{2}}^{-}\otimes0^{+}$  
				&$(1^{+}\otimes1^{+})_{1} \otimes {\frac{1}{2}}^{-}\otimes0^{+}$     &$(0^{+}\otimes1^{+})\otimes{\frac{1}{2}}^{-}\otimes0^{+}$ 
				&$(1^{+}\otimes1^{+})_{1} \otimes {\frac{1}{2}}^{-}\otimes0^{+}$
				&$(1^{+}\otimes1^{+})_{2} \otimes {\frac{1}{2}}^{-}\otimes0^{+}$   
				&$(1^{+}\otimes1^{+})\otimes {\frac{1}{2}}^{-}\otimes0^{+}$
				\\
				\hline
				$(0,1,0)$  &0.514 &-0.377 &0.368 &0.206 &-0.013 &0.881 &0.352 
				\\
				\hline
				&\multicolumn{1}{c|}{$^{2}P_{\frac{1}{2}}$	(${J^P={\frac{1}{2}}^{+}}$)}
				&\multicolumn{1}{c|}{$^{4}P_{\frac{1}{2}}$	(${J^P={\frac{1}{2}}^{+}}$)}
				&\multicolumn{2}{c|}{$^{2}P_{\frac{1}{2}}$	(${J^P={\frac{1}{2}}^{+}}$)}
				&\multicolumn{2}{c|}{$^{4}P_{\frac{1}{2}}$  (${J^P={\frac{1}{2}}^{+}}$)}
				&
				\\
				\cline{2-8} 
				$(Y, I, I_3)$
				&$(0^{+}\otimes1^{+}\otimes	{\frac{1}{2}}^{-})_{\frac{1}{2}}\otimes1^{-}$
				&$(0^{+}\otimes1^{+}\otimes	{\frac{1}{2}}^{-})_{\frac{3}{2}}\otimes1^{-}$
				&$((1^{+}\otimes1^{+})_{0}\otimes{\frac{1}{2}}^{-})_{\frac{1}{2}}\otimes1^{-}$
				&$((1^{+}\otimes1^{+})_{1}\otimes{\frac{1}{2}}^{-})_{\frac{1}{2}}\otimes1^{-}$
				&$((1^{+}\otimes1^{+})_{1}\otimes{\frac{1}{2}}^{-})_{\frac{3}{2}}\otimes1^{-}$    &$((1^{+}\otimes1^{+})_{2}\otimes{\frac{1}{2}}^{-})_{\frac{3}{2}}\otimes1^{-}$
				\\
				\hline		
				$(0,1,0)$ &-0.035&0.046& 0.260& 0.012&-0.074& 0.422
				\\
				\hline
				&\multicolumn{1}{c|}{${^2 P_{\frac{3}{2}}}$ (${J^P={\frac{3}{2}}^{+}}$)}
				&\multicolumn{1}{c|}{${^4 P_{\frac{3}{2}}}$ (${J^P={\frac{3}{2}}^{+}}$)}
				&\multicolumn{2}{c|}{${^2 P_{\frac{3}{2}}}$ (${J^P={\frac{3}{2}}^{+}}$)}
				&\multicolumn{2}{c|}{${^4 P_{\frac{3}{2}}}$ (${J^P={\frac{3}{2}}^{+}}$)}
				&\multicolumn{1}{c}{${^6 P_{\frac{3}{2}}}$ (${J^P={\frac{3}{2}}^{+}}$)}
				\\
				\cline{2-8}
				$(Y, I, I_3)$  
				&$(0^{+}\otimes1^{+}\otimes{\frac{1}{2}}^{-})_{\frac{1}{2}}\otimes1^{-}$
				&$(0^{+}\otimes1^{+}\otimes	{\frac{1}{2}}^{-})_{\frac{3}{2}}\otimes1^{-}$
				&$((1^{+}\otimes1^{+})_{0}\otimes{\frac{1}{2}}^{-})_{\frac{1}{2}}\otimes1^{-}$
				&$((1^{+}\otimes1^{+})_{1}\otimes{\frac{1}{2}}^{-})_{\frac{1}{2}}\otimes1^{-}$
				&$((1^{+}\otimes1^{+})_{1}\otimes{\frac{1}{2}}^{-})_{\frac{3}{2}}\otimes1^{-}$    &$((1^{+}\otimes1^{+})_{2}\otimes{\frac{1}{2}}^{-})_{\frac{3}{2}}\otimes1^{-}$
				&$((1^{+}\otimes1^{+})_{2}\otimes{\frac{1}{2}}^{-})_{\frac{5}{2}}\otimes1^{-}$
				\\
				\hline
				$(0,1,0)$  &0.719&0.233&-0.175&0.570&0.005&0.727&0.174
				\\		
				\hline		
				&\multicolumn{3}{c|}{${^{4}P_{\frac{5}{2}}^{+}}$ (${J^P={\frac{5}{2}}^{+}}$)}  &\multicolumn{1}{c|}{${^{6}P_{\frac{5}{2}}}$ (${J^P={\frac{5}{2}}^{+}}$)} &\multicolumn{1}{c|}{${^6P_{\frac{7}{2}}}$ (${J^P={\frac{7}{2}}^{+}}$)} 
				\\
				\cline{2-8}		
				$(Y, I, I_3)$  &$(0^{+}\otimes1^{+}\otimes
				{\frac{1}{2}}^{-})\otimes1^{-}$
				&$((1^{+}\otimes1^{+})_{1}\otimes{\frac{1}{2}}^{-})_{\frac{3}{2}}\otimes1^{-}$
				&$((1^{+}\otimes1^{+})_{2}\otimes{\frac{1}{2}}^{-})_{\frac{3}{2}}\otimes1^{-}$
				&$((1^{+}\otimes1^{+})_{2}\otimes{\frac{1}{2}}^{-})_{\frac{5}{2}}\otimes1^{-}$
				&$1^{+}\otimes1^{+}\otimes{\frac{1}{2}}^{-}\otimes1^{-}$
				\\
				\hline		
				$(0,1,0)$  &0.410&0.190&1.083&0.369&0.554
				\\	
				\hline
				\bottomrule[1pt]
				\multicolumn{8}{c}{$8_{2f}$: $\frac{1}{\sqrt2}\{ (cd)[us]{\bar c}+(cu)[ds]{\bar c} \}$ }
				\\
				\hline
				&\multicolumn{2}{c|}{$^2S_{\frac{1}{2}}$ ($J^P={\frac{1}{2}}^{-}$)}   &\multicolumn{1}{c|}{$^{4}S{{\frac{3}{2}}}$ (${J^P={\frac{3}{2}}^{-}}$)}  &\multicolumn{2}{c|}{${^{2}P_{\frac{1}{2}}}$ (${J^P={\frac{1}{2}}^{+}}$)}  &\multicolumn{1}{c|}{${^{4}P_{\frac{1}{2}}}$ (${J^P={\frac{1}{2}}^{+}}$)}
				\\
				\cline{2-8}
				$(Y, I, I_3)$  
				& ${0}^{+}\otimes0^{+}\otimes{\frac{1}{2}}^{-}	\otimes{0^{+}}$
				& ${1}^{+}\otimes0^{+}\otimes{\frac{1}{2}}^{-} \otimes{0^{+}}$  
				& ${1}^{+}\otimes0^{+}\otimes{\frac{1}{2}}^{-} \otimes{0^{+}}$  
				& $0^{+}\otimes0^{+}\otimes{\frac{1}{2}}^{-} \otimes{1^{-}}$
				& $({1}^{+}\otimes0^{+}\otimes{\frac{1}{2}}^{-})_{\frac{1}{2}} \otimes{1^{-}}$ 
				& $({1}^{+}\otimes0^{+}\otimes{\frac{1}{2}}^{-})_{\frac{3}{2}}\otimes{1^{-}}$
				\\
				\hline
				$(0,1,0)$ &-0.377&0.687&0.465&0.137&-0.224&0.256
				\\
				\hline
				&\multicolumn{2}{c|}{$^{2}P_{\frac{3}{2}}$($J^P={\frac{3}{2}}^{+}$)}
				&\multicolumn{1}{c|}{$^{4}P_{\frac{3}{2}}$($J^P={\frac{3}{2}}^{+}$)}
				&\multicolumn{1}{c|}{${^{4}P_{\frac{5}{2}}^{+}}$(${J^P={\frac{5}{2}}^{+}}$)} 
				\\
				\cline{2-8} 
				$(Y, I, I_3)$ 
				& ${0}^{+}\otimes0^{+}\otimes{\frac{1}{2}}^{-} \otimes{1^{-}}$
				& $({1}^{+}\otimes0^{+}\otimes{\frac{1}{2}}^{-} )_{\frac{1}{2}}\otimes{1^{-}}$ 
				& $({1}^{+}\otimes0^{+}\otimes{\frac{1}{2}}^{-})_{\frac{3}{2}} \otimes{1^{-}}$ 
				& ${1}^{+}\otimes0^{+}\otimes{\frac{1}{2}}^{-} \otimes{1^{-}}$
				\\
				\hline
				$(0,1,0)$ &-0.360&0.695&0.344 &0.473
				\\
				\hline
				\bottomrule[1pt]
		\end{tabular}}
	\end{center}
\end{table*}

\begin{table*}[htbp]
	\caption{The magnetic moments of the pentaquark states in the diquark-diquark-antiquark model with the wave function  $\frac{1}{\sqrt2}[({c}u)\{ds\}{\bar c}-({c}d)\{us\}{\bar c}]$ in $8_{1f}$ and $\frac{1}{\sqrt6} \{ (cd)[us]{\bar c}-(cu)[ds]{\bar c}-2(cs)[ud]{\bar c} \}$ in $8_{2f}$ with isospin $(I,I_3) = (0,0)$. The third line $J_{1}^{P_{1}}\otimes J_{2}^{P_{2}}\otimes J_{3}^{P_{3}}\otimes J_{4}^{P_{4}}$ are corresponding to the angular momentum and parity of $(cq_1)$, $(q_2q_3)$, $\bar{c}$ and orbital, respectively. The $\rho$ mode P-wave orbital excitation lies between the diquark $(cq_1)$ and diquark $(q_2q_3)$. }
	\label{qag}
	\begin{center}
		\resizebox{0.95\columnwidth}{!}{
			\begin{tabular}{c|c|c|c|c|c|c|c} \toprule[1pt]
				\hline
				\multicolumn{8}{c} {$8_{1f}$: $\frac{1}{\sqrt2}[({c}u)\{ds\}{\bar c}-({c}d)\{us\}{\bar c}]$}
				\\
				\hline
				&\multicolumn{3}{c|}{$^{2}S_{\frac{1}{2}}$($J^P={\frac{1}{2}}^{-}$)}
				&\multicolumn{3}{c|}{${^{4}S_{\frac{3}{2}}}$(${J^P={\frac{3}{2}}^{-}}$)}
				&\multicolumn{1}{c}{$^{6}S_{\frac{5}{2}}$ ($J^P={\frac{5}{2}}^{-}$)}
				\\
				\cline{2-8}
				$(Y, I, I_3)$   
				& $0^{+}\otimes1^{+} \otimes{\frac{1}{2}}^{-}\otimes0^{+}$
				&$(1^{+}\otimes1^{+})_{0} \otimes {\frac{1}{2}}^{-}\otimes0^{+}$  
				&$(1^{+}\otimes1^{+})_{1} \otimes {\frac{1}{2}}^{-}\otimes0^{+}$     &$(0^{+}\otimes1^{+})\otimes{\frac{1}{2}}^{-}\otimes0^{+}$ 
				&$(1^{+}\otimes1^{+})_{1} \otimes {\frac{1}{2}}^{-}\otimes0^{+}$
				&$(1^{+}\otimes1^{+})_{2} \otimes {\frac{1}{2}}^{-}\otimes0^{+}$   
				&$(1^{+}\otimes1^{+})\otimes {\frac{1}{2}}^{-}\otimes0^{+}$
				\\
				\hline
				$(0,0,0)$  &0.050 &-0.377 &0.368 &-0.490 &-0.013 &0.881 &0.352
				\\
				\hline
				&\multicolumn{1}{c|}{$^{2}P_{\frac{1}{2}}$	(${J^P={\frac{1}{2}}^{+}}$)}
				&\multicolumn{1}{c|}{$^{4}P_{\frac{1}{2}}$	(${J^P={\frac{1}{2}}^{+}}$)}
				&\multicolumn{2}{c|}{$^{2}P_{\frac{1}{2}}$	(${J^P={\frac{1}{2}}^{+}}$)}
				&\multicolumn{2}{c|}{$^{4}P_{\frac{1}{2}}$  (${J^P={\frac{1}{2}}^{+}}$)}
				&
				\\
				\cline{2-8} 
				$(Y, I, I_3)$
				&$(0^{+}\otimes1^{+}\otimes	{\frac{1}{2}}^{-})_{\frac{1}{2}}\otimes1^{-}$
				&$(0^{+}\otimes1^{+}\otimes	{\frac{1}{2}}^{-})_{\frac{3}{2}}\otimes1^{-}$
				&$((1^{+}\otimes1^{+})_{0}\otimes{\frac{1}{2}}^{-})_{\frac{1}{2}}\otimes1^{-}$
				&$((1^{+}\otimes1^{+})_{1}\otimes{\frac{1}{2}}^{-})_{\frac{1}{2}}\otimes1^{-}$
				&$((1^{+}\otimes1^{+})_{1}\otimes{\frac{1}{2}}^{-})_{\frac{3}{2}}\otimes1^{-}$    &$((1^{+}\otimes1^{+})_{2}\otimes{\frac{1}{2}}^{-})_{\frac{3}{2}}\otimes1^{-}$
				\\
				\hline		
				$(0,0,0)$ &0.013 &-0.287&0.150&-0.098&-0.019& 0.478
				\\
				\hline
				&\multicolumn{1}{c|}{${^2 P_{\frac{3}{2}}}$ (${J^P={\frac{3}{2}}^{+}}$)}
				&\multicolumn{1}{c|}{${^4 P_{\frac{3}{2}}}$ (${J^P={\frac{3}{2}}^{+}}$)}
				&\multicolumn{2}{c|}{${^2 P_{\frac{3}{2}}}$ (${J^P={\frac{3}{2}}^{+}}$)}
				&\multicolumn{2}{c|}{${^4 P_{\frac{3}{2}}}$ (${J^P={\frac{3}{2}}^{+}}$)}
				&\multicolumn{1}{c}{${^6 P_{\frac{3}{2}}}$ (${J^P={\frac{3}{2}}^{+}}$)}
				\\
				\cline{2-8}
				$(Y, I, I_3)$  
				&$(0^{+}\otimes1^{+}\otimes{\frac{1}{2}}^{-})_{\frac{1}{2}}\otimes1^{-}$
				&$(0^{+}\otimes1^{+}\otimes	{\frac{1}{2}}^{-})_{\frac{3}{2}}\otimes1^{-}$
				&$((1^{+}\otimes1^{+})_{0}\otimes{\frac{1}{2}}^{-})_{\frac{1}{2}}\otimes1^{-}$
				&$((1^{+}\otimes1^{+})_{1}\otimes{\frac{1}{2}}^{-})_{\frac{1}{2}}\otimes1^{-}$
				&$((1^{+}\otimes1^{+})_{1}\otimes{\frac{1}{2}}^{-})_{\frac{3}{2}}\otimes1^{-}$    &$((1^{+}\otimes1^{+})_{2}\otimes{\frac{1}{2}}^{-})_{\frac{3}{2}}\otimes1^{-}$
				&$((1^{+}\otimes1^{+})_{2}\otimes{\frac{1}{2}}^{-})_{\frac{5}{2}}\otimes1^{-}$
				\\
				\hline
				$(0, 0,0)$  &0.094&-0.342&-0.340&0.405&0.197&0.661&0.273
				\\		
				\hline		
				&\multicolumn{3}{c|}{${^{4}P_{\frac{5}{2}}^{+}}$ (${J^P={\frac{5}{2}}^{+}}$)}  &\multicolumn{1}{c|}{${^{6}P_{\frac{5}{2}}}$ (${J^P={\frac{5}{2}}^{+}}$)} &\multicolumn{1}{c|}{${^6P_{\frac{7}{2}}}$ (${J^P={\frac{7}{2}}^{+}}$)} 
				\\
				\cline{2-8}		
				$(Y, I, I_3)$  &$(0^{+}\otimes1^{+}\otimes
				{\frac{1}{2}}^{-})\otimes1^{-}$
				&$((1^{+}\otimes1^{+})_{1}\otimes{\frac{1}{2}}^{-})_{\frac{3}{2}}\otimes1^{-}$
				&$((1^{+}\otimes1^{+})_{2}\otimes{\frac{1}{2}}^{-})_{\frac{3}{2}}\otimes1^{-}$
				&$((1^{+}\otimes1^{+})_{2}\otimes{\frac{1}{2}}^{-})_{\frac{5}{2}}\otimes1^{-}$
				&$1^{+}\otimes1^{+}\otimes{\frac{1}{2}}^{-}\otimes1^{-}$
				\\
				\hline		
				$(0,0,0)$  &-0.446&0.024&0.918&0.322&0.388
				\\	
				\hline
				\bottomrule[1pt]
				\multicolumn{8}{c}{ $8_{2f}$: $\frac{1}{\sqrt6} \{ (cd)[us]{\bar c}-(cu)[ds]{\bar c}-2(cs)[ud]{\bar c} \}$ }
				\\
				\hline
				&\multicolumn{2}{c|}{$^2S_{\frac{1}{2}}$ ($J^P={\frac{1}{2}}^{-}$)}   &\multicolumn{1}{c|}{$^{4}S{{\frac{3}{2}}}$ (${J^P={\frac{3}{2}}^{-}}$)}  &\multicolumn{2}{c|}{${^{2}P_{\frac{1}{2}}}$ (${J^P={\frac{1}{2}}^{+}}$)}  &\multicolumn{1}{c|}{${^{4}P_{\frac{1}{2}}}$ (${J^P={\frac{1}{2}}^{+}}$)}
				\\
				\cline{2-8}
				$(Y, I, I_3)$  
				& ${0}^{+}\otimes0^{+}\otimes{\frac{1}{2}}^{-}	\otimes{0^{+}}$
				& ${1}^{+}\otimes0^{+}\otimes{\frac{1}{2}}^{-} \otimes{0^{+}}$  
				& ${1}^{+}\otimes0^{+}\otimes{\frac{1}{2}}^{-} \otimes{0^{+}}$  
				& $0^{+}\otimes0^{+}\otimes{\frac{1}{2}}^{-} \otimes{1^{-}}$
				& $({1}^{+}\otimes0^{+}\otimes{\frac{1}{2}}^{-})_{\frac{1}{2}} \otimes{1^{-}}$ 
				& $({1}^{+}\otimes0^{+}\otimes{\frac{1}{2}}^{-})_{\frac{3}{2}}\otimes{1^{-}}$
				\\
				\hline
				$(0,0,0)$ &-0.377&0.223&-0.231&0.292&0.091&-0.211
				\\
				\hline
				&\multicolumn{2}{c|}{$^{2}P_{\frac{3}{2}}$($J^P={\frac{3}{2}}^{+}$)}
				&\multicolumn{1}{c|}{$^{4}P_{\frac{3}{2}}$($J^P={\frac{3}{2}}^{+}$)}
				&\multicolumn{1}{c|}{${^{4}P_{\frac{5}{2}}^{+}}$(${J^P={\frac{5}{2}}^{+}}$)} 
				\\
				\cline{2-8} 
				$(Y, I, I_3)$ 
				& ${0}^{+}\otimes0^{+}\otimes{\frac{1}{2}}^{-} \otimes{1^{-}}$
				& $({1}^{+}\otimes0^{+}\otimes{\frac{1}{2}}^{-} )_{\frac{1}{2}}\otimes{1^{-}}$ 
				& $({1}^{+}\otimes0^{+}\otimes{\frac{1}{2}}^{-})_{\frac{3}{2}} \otimes{1^{-}}$ 
				& ${1}^{+}\otimes0^{+}\otimes{\frac{1}{2}}^{-} \otimes{1^{-}}$
				\\
				\hline
				$(0,0,0)$ &-0.126&0.470&-0.070 &0.016
				\\
				\hline
				\bottomrule[1pt]
		\end{tabular}}
	\end{center}
\end{table*}

\begin{table*}[htbp]
	\caption{The magnetic moments of the pentaquark states in the diquark-diquark-antiquark model with the wave function  $\frac{1}{\sqrt6}[({c}d)\{us\}{\bar c}+({c}u)\{ds\}{\bar c}]-\sqrt{\frac{2}{3}}({c}s)\{ud\}{\bar c}$ in $8_{1f}$ and $\frac{1}{\sqrt2}\{ (cd)[us]{\bar c}+(cu)[ds]{\bar c} \}$ in $8_{2f}$ with isospin $(I,I_3) = (1,0)$. The third line $J_{1}^{P_{1}}\otimes J_{2}^{P_{2}}\otimes J_{3}^{P_{3}}\otimes J_{4}^{P_{4}}$ are corresponding to the angular momentum and parity of $(cq_1)$, $(q_2q_3)$, $\bar{c}$ and orbital, respectively. The $\lambda$ mode P-wave orbital excitation lies between the $\bar{c}$ and the center of mass system of the $(cq_1)$ and $(q_2q_3)$.The unit is the magnetic moments of the proton.}
	\label{abc}
	\begin{center}
		\resizebox{0.95\columnwidth}{!}{
			\begin{tabular}{c|c|c|c|c|c|c|c} \toprule[1pt]
				\hline
				\multicolumn{8}{c} {$8_{1f}$: $\frac{1}{\sqrt6}[({c}d)\{us\}{\bar c}+({c}u)\{ds\}{\bar c}]-\sqrt{\frac{2}{3}}({c}s)\{ud\}{\bar c}$}
				\\
				\hline
				&\multicolumn{3}{c|}{$^{2}S_{\frac{1}{2}}$($J^P={\frac{1}{2}}^{-}$)}
				&\multicolumn{3}{c|}{${^{4}S_{\frac{3}{2}}}$(${J^P={\frac{3}{2}}^{-}}$)}
				&\multicolumn{1}{c}{$^{6}S_{\frac{5}{2}}$ ($J^P={\frac{5}{2}}^{-}$)}
				\\
				\cline{2-8}
				$(Y, I, I_3)$   
				& $0^{+}\otimes1^{+} \otimes{\frac{1}{2}}^{-}\otimes0^{+}$
				&$(1^{+}\otimes1^{+})_{0} \otimes {\frac{1}{2}}^{-}\otimes0^{+}$  
				&$(1^{+}\otimes1^{+})_{1} \otimes {\frac{1}{2}}^{-}\otimes0^{+}$     &$(0^{+}\otimes1^{+})\otimes{\frac{1}{2}}^{-}\otimes0^{+}$ 
				&$(1^{+}\otimes1^{+})_{1} \otimes {\frac{1}{2}}^{-}\otimes0^{+}$
				&$(1^{+}\otimes1^{+})_{2} \otimes {\frac{1}{2}}^{-}\otimes0^{+}$   
				&$(1^{+}\otimes1^{+})\otimes {\frac{1}{2}}^{-}\otimes0^{+}$
				\\
				\hline
				$(0,1,0)$  &0.514 &-0.377 &0.368 &0.206 &-0.013 &0.881 &0.352 
				\\
				\hline
				&\multicolumn{1}{c|}{$^{2}P_{\frac{1}{2}}$	(${J^P={\frac{1}{2}}^{+}}$)}
				&\multicolumn{1}{c|}{$^{4}P_{\frac{1}{2}}$	(${J^P={\frac{1}{2}}^{+}}$)}
				&\multicolumn{2}{c|}{$^{2}P_{\frac{1}{2}}$	(${J^P={\frac{1}{2}}^{+}}$)}
				&\multicolumn{2}{c|}{$^{4}P_{\frac{1}{2}}$  (${J^P={\frac{1}{2}}^{+}}$)}
				&
				\\
				\cline{2-8} 
				$(Y, I, I_3)$
				&$(0^{+}\otimes1^{+}\otimes	{\frac{1}{2}}^{-})_{\frac{1}{2}}\otimes1^{-}$
				&$(0^{+}\otimes1^{+}\otimes	{\frac{1}{2}}^{-})_{\frac{3}{2}}\otimes1^{-}$
				&$((1^{+}\otimes1^{+})_{0}\otimes{\frac{1}{2}}^{-})_{\frac{1}{2}}\otimes1^{-}$
				&$((1^{+}\otimes1^{+})_{1}\otimes{\frac{1}{2}}^{-})_{\frac{1}{2}}\otimes1^{-}$
				&$((1^{+}\otimes1^{+})_{1}\otimes{\frac{1}{2}}^{-})_{\frac{3}{2}}\otimes1^{-}$    &$((1^{+}\otimes1^{+})_{2}\otimes{\frac{1}{2}}^{-})_{\frac{3}{2}}\otimes1^{-}$
				\\
				\hline		
				$(0,1,0)$ &0.217 &-0.080&0.507&0.259&-0.198& 0.299
				\\
				\hline
				&\multicolumn{1}{c|}{${^2 P_{\frac{3}{2}}}$ (${J^P={\frac{3}{2}}^{+}}$)}
				&\multicolumn{1}{c|}{${^4 P_{\frac{3}{2}}}$ (${J^P={\frac{3}{2}}^{+}}$)}
				&\multicolumn{2}{c|}{${^2 P_{\frac{3}{2}}}$ (${J^P={\frac{3}{2}}^{+}}$)}
				&\multicolumn{2}{c|}{${^4 P_{\frac{3}{2}}}$ (${J^P={\frac{3}{2}}^{+}}$)}
				&\multicolumn{1}{c}{${^6 P_{\frac{3}{2}}}$ (${J^P={\frac{3}{2}}^{+}}$)}
				\\
				\cline{2-8}
				$(Y, I, I_3)$  
				&$(0^{+}\otimes1^{+}\otimes{\frac{1}{2}}^{-})_{\frac{1}{2}}\otimes1^{-}$
				&$(0^{+}\otimes1^{+}\otimes	{\frac{1}{2}}^{-})_{\frac{3}{2}}\otimes1^{-}$
				&$((1^{+}\otimes1^{+})_{0}\otimes{\frac{1}{2}}^{-})_{\frac{1}{2}}\otimes1^{-}$
				&$((1^{+}\otimes1^{+})_{1}\otimes{\frac{1}{2}}^{-})_{\frac{1}{2}}\otimes1^{-}$
				&$((1^{+}\otimes1^{+})_{1}\otimes{\frac{1}{2}}^{-})_{\frac{3}{2}}\otimes1^{-}$    &$((1^{+}\otimes1^{+})_{2}\otimes{\frac{1}{2}}^{-})_{\frac{3}{2}}\otimes1^{-}$
				&$((1^{+}\otimes1^{+})_{2}\otimes{\frac{1}{2}}^{-})_{\frac{5}{2}}\otimes1^{-}$
				\\
				\hline
				$(0,1,0)$  &1.096&0.384&0.196&0.941&0.220&0.875&-0.048
				\\		
				\hline		
				&\multicolumn{3}{c|}{${^{4}P_{\frac{5}{2}}^{+}}$ (${J^P={\frac{5}{2}}^{+}}$)}  &\multicolumn{1}{c|}{${^{6}P_{\frac{5}{2}}}$ (${J^P={\frac{5}{2}}^{+}}$)} &\multicolumn{1}{c|}{${^6P_{\frac{7}{2}}}$ (${J^P={\frac{7}{2}}^{+}}$)} 
				\\
				\cline{2-8}		
				$(Y, I, I_3)$  &$(0^{+}\otimes1^{+}\otimes
				{\frac{1}{2}}^{-})\otimes1^{-}$
				&$((1^{+}\otimes1^{+})_{1}\otimes{\frac{1}{2}}^{-})_{\frac{3}{2}}\otimes1^{-}$
				&$((1^{+}\otimes1^{+})_{2}\otimes{\frac{1}{2}}^{-})_{\frac{3}{2}}\otimes1^{-}$
				&$((1^{+}\otimes1^{+})_{2}\otimes{\frac{1}{2}}^{-})_{\frac{5}{2}}\otimes1^{-}$
				&$1^{+}\otimes1^{+}\otimes{\frac{1}{2}}^{-}\otimes1^{-}$
				\\
				\hline		
				$(0,1,0)$  &0.788&0.560&1.454&0.475&0.924
				\\	
				\hline
				\bottomrule[1pt]
				\multicolumn{8}{c}{$8_{2f}$: $\frac{1}{\sqrt2}\{ (cd)[us]{\bar c}+(cu)[ds]{\bar c} \}$ }
				\\
				\hline
				&\multicolumn{2}{c|}{$^2S_{\frac{1}{2}}$ ($J^P={\frac{1}{2}}^{-}$)}   &\multicolumn{1}{c|}{$^{4}S{{\frac{3}{2}}}$ (${J^P={\frac{3}{2}}^{-}}$)}  &\multicolumn{2}{c|}{${^{2}P_{\frac{1}{2}}}$ (${J^P={\frac{1}{2}}^{+}}$)}  &\multicolumn{1}{c|}{${^{4}P_{\frac{1}{2}}}$ (${J^P={\frac{1}{2}}^{+}}$)}
				\\
				\cline{2-8}
				$(Y, I, I_3)$  
				& ${0}^{+}\otimes0^{+}\otimes{\frac{1}{2}}^{-}	\otimes{0^{+}}$
				& ${1}^{+}\otimes0^{+}\otimes{\frac{1}{2}}^{-} \otimes{0^{+}}$  
				& ${1}^{+}\otimes0^{+}\otimes{\frac{1}{2}}^{-} \otimes{0^{+}}$  
				& $0^{+}\otimes0^{+}\otimes{\frac{1}{2}}^{-} \otimes{1^{-}}$
				& $({1}^{+}\otimes0^{+}\otimes{\frac{1}{2}}^{-})_{\frac{1}{2}} \otimes{1^{-}}$ 
				& $({1}^{+}\otimes0^{+}\otimes{\frac{1}{2}}^{-})_{\frac{3}{2}}\otimes{1^{-}}$
				\\
				\hline
				$(0,1,0)$ &-0.377&0.687&0.465&0.525&0.164&0.062
				\\
				\hline
				&\multicolumn{2}{c|}{$^{2}P_{\frac{3}{2}}$($J^P={\frac{3}{2}}^{+}$)}
				&\multicolumn{1}{c|}{$^{4}P_{\frac{3}{2}}$($J^P={\frac{3}{2}}^{+}$)}
				&\multicolumn{1}{c|}{${^{4}P_{\frac{5}{2}}^{+}}$(${J^P={\frac{5}{2}}^{+}}$)} 
				\\
				\cline{2-8} 
				$(Y, I, I_3)$ 
				& ${0}^{+}\otimes0^{+}\otimes{\frac{1}{2}}^{-} \otimes{1^{-}}$
				& $({1}^{+}\otimes0^{+}\otimes{\frac{1}{2}}^{-} )_{\frac{1}{2}}\otimes{1^{-}}$ 
				& $({1}^{+}\otimes0^{+}\otimes{\frac{1}{2}}^{-})_{\frac{3}{2}} \otimes{1^{-}}$ 
				& ${1}^{+}\otimes0^{+}\otimes{\frac{1}{2}}^{-} \otimes{1^{-}}$
				\\
				\hline
				$(0,1,0)$ &0.223&1.277&0.577 &1.055 
				\\
				\hline
				\bottomrule[1pt]
		\end{tabular}}
	\end{center}
\end{table*}
\begin{table*}[htbp]
	\caption{The magnetic moments of the pentaquark states in the diquark-diquark-antiquark model with the wave function  $\frac{1}{\sqrt2}[({c}u)\{ds\}{\bar c}-({c}d)\{us\}{\bar c}]$ in $8_{1f}$ and $\frac{1}{\sqrt6} \{ (cd)[us]{\bar c}-(cu)[ds]{\bar c}-2(cs)[ud]{\bar c} \}$ in $8_{2f}$ with isospin $(I,I_3) = (0,0)$. The third line $J_{1}^{P_{1}}\otimes J_{2}^{P_{2}}\otimes J_{3}^{P_{3}}\otimes J_{4}^{P_{4}}$ are corresponding to the angular momentum and parity of $(cq_1)$, $(q_2q_3)$, $\bar{c}$ and orbital, respectively. The $\lambda$ mode P-wave orbital excitation lies between the $\bar{c}$ and the center of mass system of the $(cq_1)$ and $(q_2q_3)$.}
	\label{def}
	\begin{center}
		\resizebox{0.95\columnwidth}{!}{
			\begin{tabular}{c|c|c|c|c|c|c|c} \toprule[1pt]
				\hline
				\multicolumn{8}{c} {$8_{1f}$: $\frac{1}{\sqrt2}[({c}u)\{ds\}{\bar c}-({c}d)\{us\}{\bar c}]$}
				\\
				\hline
				&\multicolumn{3}{c|}{$^{2}S_{\frac{1}{2}}$($J^P={\frac{1}{2}}^{-}$)}
				&\multicolumn{3}{c|}{${^{4}S_{\frac{3}{2}}}$(${J^P={\frac{3}{2}}^{-}}$)}
				&\multicolumn{1}{c}{$^{6}S_{\frac{5}{2}}$ ($J^P={\frac{5}{2}}^{-}$)}
				\\
				\cline{2-8}
				$(Y, I, I_3)$   
				& $0^{+}\otimes1^{+} \otimes{\frac{1}{2}}^{-}\otimes0^{+}$
				&$(1^{+}\otimes1^{+})_{0} \otimes {\frac{1}{2}}^{-}\otimes0^{+}$  
				&$(1^{+}\otimes1^{+})_{1} \otimes {\frac{1}{2}}^{-}\otimes0^{+}$     &$(0^{+}\otimes1^{+})\otimes{\frac{1}{2}}^{-}\otimes0^{+}$ 
				&$(1^{+}\otimes1^{+})_{1} \otimes {\frac{1}{2}}^{-}\otimes0^{+}$
				&$(1^{+}\otimes1^{+})_{2} \otimes {\frac{1}{2}}^{-}\otimes0^{+}$   
				&$(1^{+}\otimes1^{+})\otimes {\frac{1}{2}}^{-}\otimes0^{+}$
				\\
				\hline
				$(0,0,0)$  &0.050 &-0.377 &0.368 &-0.490 &-0.013 &0.881 &0.352
				\\
				\hline
				&\multicolumn{1}{c|}{$^{2}P_{\frac{1}{2}}$	(${J^P={\frac{1}{2}}^{+}}$)}
				&\multicolumn{1}{c|}{$^{4}P_{\frac{1}{2}}$	(${J^P={\frac{1}{2}}^{+}}$)}
				&\multicolumn{2}{c|}{$^{2}P_{\frac{1}{2}}$	(${J^P={\frac{1}{2}}^{+}}$)}
				&\multicolumn{2}{c|}{$^{4}P_{\frac{1}{2}}$  (${J^P={\frac{1}{2}}^{+}}$)}
				&
				\\
				\cline{2-8} 
				$(Y, I, I_3)$
				&$(0^{+}\otimes1^{+}\otimes	{\frac{1}{2}}^{-})_{\frac{1}{2}}\otimes1^{-}$
				&$(0^{+}\otimes1^{+}\otimes	{\frac{1}{2}}^{-})_{\frac{3}{2}}\otimes1^{-}$
				&$((1^{+}\otimes1^{+})_{0}\otimes{\frac{1}{2}}^{-})_{\frac{1}{2}}\otimes1^{-}$
				&$((1^{+}\otimes1^{+})_{1}\otimes{\frac{1}{2}}^{-})_{\frac{1}{2}}\otimes1^{-}$
				&$((1^{+}\otimes1^{+})_{1}\otimes{\frac{1}{2}}^{-})_{\frac{3}{2}}\otimes1^{-}$    &$((1^{+}\otimes1^{+})_{2}\otimes{\frac{1}{2}}^{-})_{\frac{3}{2}}\otimes1^{-}$
				\\
				\hline		
				$(0,0,0)$ &0.334 &-0.448&0.469&0.221&-0.179& 0.318
				\\
				\hline
				&\multicolumn{1}{c|}{${^2 P_{\frac{3}{2}}}$ (${J^P={\frac{3}{2}}^{+}}$)}
				&\multicolumn{1}{c|}{${^4 P_{\frac{3}{2}}}$ (${J^P={\frac{3}{2}}^{+}}$)}
				&\multicolumn{2}{c|}{${^2 P_{\frac{3}{2}}}$ (${J^P={\frac{3}{2}}^{+}}$)}
				&\multicolumn{2}{c|}{${^4 P_{\frac{3}{2}}}$ (${J^P={\frac{3}{2}}^{+}}$)}
				&\multicolumn{1}{c}{${^6 P_{\frac{3}{2}}}$ (${J^P={\frac{3}{2}}^{+}}$)}
				\\
				\cline{2-8}
				$(Y, I, I_3)$  
				&$(0^{+}\otimes1^{+}\otimes{\frac{1}{2}}^{-})_{\frac{1}{2}}\otimes1^{-}$
				&$(0^{+}\otimes1^{+}\otimes	{\frac{1}{2}}^{-})_{\frac{3}{2}}\otimes1^{-}$
				&$((1^{+}\otimes1^{+})_{0}\otimes{\frac{1}{2}}^{-})_{\frac{1}{2}}\otimes1^{-}$
				&$((1^{+}\otimes1^{+})_{1}\otimes{\frac{1}{2}}^{-})_{\frac{1}{2}}\otimes1^{-}$
				&$((1^{+}\otimes1^{+})_{1}\otimes{\frac{1}{2}}^{-})_{\frac{3}{2}}\otimes1^{-}$    &$((1^{+}\otimes1^{+})_{2}\otimes{\frac{1}{2}}^{-})_{\frac{3}{2}}\otimes1^{-}$
				&$((1^{+}\otimes1^{+})_{2}\otimes{\frac{1}{2}}^{-})_{\frac{5}{2}}\otimes1^{-}$
				\\
				\hline
				$(0, 0,0)$  &0.575&-0.150&0.139&0.884&0.072&0.853&-0.014
				\\		
				\hline		
				&\multicolumn{3}{c|}{${^{4}P_{\frac{5}{2}}^{+}}$ (${J^P={\frac{5}{2}}^{+}}$)}  &\multicolumn{1}{c|}{${^{6}P_{\frac{5}{2}}}$ (${J^P={\frac{5}{2}}^{+}}$)} &\multicolumn{1}{c|}{${^6P_{\frac{7}{2}}}$ (${J^P={\frac{7}{2}}^{+}}$)} 
				\\
				\cline{2-8}		
				$(Y, I, I_3)$  &$(0^{+}\otimes1^{+}\otimes
				{\frac{1}{2}}^{-})\otimes1^{-}$
				&$((1^{+}\otimes1^{+})_{1}\otimes{\frac{1}{2}}^{-})_{\frac{3}{2}}\otimes1^{-}$
				&$((1^{+}\otimes1^{+})_{2}\otimes{\frac{1}{2}}^{-})_{\frac{3}{2}}\otimes1^{-}$
				&$((1^{+}\otimes1^{+})_{2}\otimes{\frac{1}{2}}^{-})_{\frac{5}{2}}\otimes1^{-}$
				&$1^{+}\otimes1^{+}\otimes{\frac{1}{2}}^{-}\otimes1^{-}$
				\\
				\hline		
				$(0,0,0)$  &0.035&0.503&1.397&0.459&0.867
				\\	
				\hline
				\bottomrule[1pt]
				\multicolumn{8}{c}{ $8_{2f}$: $\frac{1}{\sqrt6} \{ (cd)[us]{\bar c}-(cu)[ds]{\bar c}-2(cs)[ud]{\bar c} \}$ }
				\\
				\hline
				&\multicolumn{2}{c|}{$^2S_{\frac{1}{2}}$ ($J^P={\frac{1}{2}}^{-}$)}   &\multicolumn{1}{c|}{$^{4}S{{\frac{3}{2}}}$ (${J^P={\frac{3}{2}}^{-}}$)}  &\multicolumn{2}{c|}{${^{2}P_{\frac{1}{2}}}$ (${J^P={\frac{1}{2}}^{+}}$)}  &\multicolumn{1}{c|}{${^{4}P_{\frac{1}{2}}}$ (${J^P={\frac{1}{2}}^{+}}$)}
				\\
				\cline{2-8}
				$(Y, I, I_3)$  
				& ${0}^{+}\otimes0^{+}\otimes{\frac{1}{2}}^{-}	\otimes{0^{+}}$
				& ${1}^{+}\otimes0^{+}\otimes{\frac{1}{2}}^{-} \otimes{0^{+}}$  
				& ${1}^{+}\otimes0^{+}\otimes{\frac{1}{2}}^{-} \otimes{0^{+}}$  
				& $0^{+}\otimes0^{+}\otimes{\frac{1}{2}}^{-} \otimes{1^{-}}$
				& $({1}^{+}\otimes0^{+}\otimes{\frac{1}{2}}^{-})_{\frac{1}{2}} \otimes{1^{-}}$ 
				& $({1}^{+}\otimes0^{+}\otimes{\frac{1}{2}}^{-})_{\frac{3}{2}}\otimes{1^{-}}$
				\\
				\hline
				$(0,0,0)$ &-0.377&0.223&-0.231&0.616&0.410&-0.370
				\\
				\hline
				&\multicolumn{2}{c|}{$^{2}P_{\frac{3}{2}}$($J^P={\frac{3}{2}}^{+}$)}
				&\multicolumn{1}{c|}{$^{4}P_{\frac{3}{2}}$($J^P={\frac{3}{2}}^{+}$)}
				&\multicolumn{1}{c|}{${^{4}P_{\frac{5}{2}}^{+}}$(${J^P={\frac{5}{2}}^{+}}$)} 
				\\
				\cline{2-8} 
				$(Y, I, I_3)$ 
				& ${0}^{+}\otimes0^{+}\otimes{\frac{1}{2}}^{-} \otimes{1^{-}}$
				& $({1}^{+}\otimes0^{+}\otimes{\frac{1}{2}}^{-} )_{\frac{1}{2}}\otimes{1^{-}}$ 
				& $({1}^{+}\otimes0^{+}\otimes{\frac{1}{2}}^{-})_{\frac{3}{2}} \otimes{1^{-}}$ 
				& ${1}^{+}\otimes0^{+}\otimes{\frac{1}{2}}^{-} \otimes{1^{-}}$
				\\
				\hline
				$(0,0,0)$ &0.359&0.949&0.121 &0.495 
				\\
				\hline
				\bottomrule[1pt]
		\end{tabular}}
	\end{center}
\end{table*}

	\subsection{Magnetic moments of the diquark-triquark model with the configuration$(cq_1)(\bar{c}q_2q_3)$}
Considering the diquark-triquark model, the total magnetic moments formula is:
\begin{eqnarray}
	\hat{\mu}  = \ \hat{\mu}_{\mathcal{D}}+\hat{\mu}_{\mathcal{T}}+\hat{\mu}_{l}. 
\end{eqnarray}
where the $l$ is the orbital excitation between the diquark and triquark. The magnetic moments formula of the pentaquark, $(cq_1)(\bar{c}q_2q_3)$ in the diquark-triquark model is 
\begin{eqnarray}
	\mu  
	&=& \langle\ \psi\ |\ \hat{\mu}_{\mathcal{D}}+\hat{\mu}_{\mathcal{T}}+\hat{\mu}_{l}\ |\ \psi\  \rangle\nonumber\\
	&=&
	\sum_{S_z,l_z}\ \langle\ SS_z,ll_z|JJ_z\ \rangle^{2}  \left \{ \mu_{l} l_z 	+
	\sum_{\widetilde{S}_{\mathcal{D}},\widetilde{S}_{\mathcal{T}}}\ \langle\ S_\mathcal{D} \widetilde{S}_{\mathcal{D}},S_\mathcal{T} \widetilde{S}_{\mathcal{T}}|SS_z\ \rangle^{2} \Bigg [
	\widetilde{S}_{\mathcal{D}}\bigg(\mu_{c} + \mu_{q_1}\bigg )\nonumber\right.\\
	&+&\left.
	\sum_{\widetilde{S}_{\bar{c}}}\ \langle\ S_{\bar{c}} \widetilde{S}_{\bar{c}},S_{r} \widetilde{S}_{\mathcal{T}}-\widetilde{S}_{\bar{c}}|S_{\mathcal{T}} \widetilde{S}_{\mathcal{T}}\rangle^{2}\bigg(g\mu_{\bar{c}}\widetilde{S}_{\bar{c}}+(\widetilde{S}_{\mathcal{T}}-\widetilde{S}_{\bar{c}})(\mu_{q_2}+\mu_{q_3})\bigg )
	\Bigg ]\right \}.           
\end{eqnarray}
where $S_{\mathcal{D}}$, $S_{\mathcal{T}}$ and $S_{r}$ represent the  diquark, triquark and the light diquark spin inside the triquark, respectively. The triquark's masses roughly use the sum of the mass of the corresponding diquark and the antiquark. The numerical results with isospin $(I,I_3) = (1,0)$ and $(I,I_3) = (0,0)$ are reported in Table \ref{ooo} and \ref{lll}, respectively.
\begin{table*}[htbp]
	\caption{The magnetic moments of the pentaquark states in the diquark-triquark model with the wave function  $\frac{1}{\sqrt6}[({c}d)(\bar c\{us\})+({c}u)(\bar c\{ds\})]-\sqrt{\frac{2}{3}}({c}s)(\bar c\{ud\})$ in $8_{1f}$ and  $\frac{1}{\sqrt2}\{ ({c}d)(\bar c[us])+({c}u)(\bar c[ds]) \}$ in $8_{2f}$ with isospin $(I,I_3) = (1,0)$. They are in $8_{1f}$ representation from $6_f \otimes 3_f = 10_f \oplus 8_{1f}$ and $8_{2f}$ representation from $\bar{3}_f \otimes 3_f = 1_f \oplus8_{2f}$, respectively. The third line $J_{1}^{P_{1}}\otimes J_{2}^{P_{2}}\otimes J_{3}^{P_{3}}$ are corresponding to the angular momentum and parity of triquark, diquark and orbital, respectively. The unit is the magnetic moments of the proton.}
	\label{ooo}
	\begin{center}
		\resizebox{0.90\columnwidth}{!}{
			\begin{tabular}{c|c|c|c|c|c|c|c} \toprule[1pt]	
				\hline
				\multicolumn{8}{c}{$8_{1f}$: $\frac{1}{\sqrt6}[({c}d)(\bar c\{us\})+({c}u)(\bar c\{ds\})]-\sqrt{\frac{2}{3}}({c}s)(\bar c\{ud\})$}
				\\
				\hline
				&\multicolumn{3}{c|}{$^{2}S_{\frac{1}{2}}$	($J^P={\frac{1}{2}}^{-}$)}
				&\multicolumn{3}{c|}{${^{4}S_{\frac{3}{2}}}$(${J^P={\frac{3}{2}}^{-}}$)}
				&\multicolumn{1}{c}{$^{6}S_{\frac{5}{2}}^{-}$($J^P={\frac{5}{2}}^{-}$)} 
				\\
				\cline{2-8} 
				$(Y, I, I_3)$ 
				& ${\frac{1}{2}}^{-}\otimes0^{+}\otimes0^{+}$
				& ${\frac{1}{2}}^{-}\otimes1^{+}\otimes0^{+}$  
				& ${\frac{3}{2}}^{-}\otimes1^{+}\otimes0^{+}$      
				& ${\frac{1}{2}}^{-}\otimes1^{+}\otimes0^{+}$ 
				& ${\frac{3}{2}}^{-}\otimes0^{+}\otimes0^{+}$
				& ${\frac{3}{2}}^{-}\otimes1^{+}\otimes0^{+}$  
				& ${\frac{3}{2}}^{-}\otimes1^{+}\otimes0^{+}$
				\\
				\hline		
				$(0,1,0)$ &0.522 &-0.078 &0.051&0.666&0.178&0.188&0.352
				\\
				\hline		
				&\multicolumn{3}{c|}{${^{2}P_{\frac{1}{2}}}$ (${J^P={\frac{1}{2}}^{+}}$)} &\multicolumn{3}{c|}{${^{4}P_{\frac{1}{2}}}$ (${J^P={\frac{1}{2}}^{+}}$)}  
				&
				\\
				\cline{2-8}
				$(Y, I, I_3)$ 
				& ${\frac{1}{2}}^{-}\otimes0^{+}\otimes1^{-}$  
				& $[{\frac{1}{2}}^{-}\otimes1^{+}]_{\frac{1}{2}}\otimes1^{-}$      
				& $[{\frac{3}{2}}^{-}\otimes1^{+}]_{\frac{1}{2}}\otimes1^{-}$  
				& ${\frac{3}{2}}^{-}\otimes0^{+}\otimes1^{-}$    
				& $[{\frac{1}{2}}^{-}\otimes1^{+}]_{\frac{3}{2}}\otimes1^{-}$  
				& $[{\frac{3}{2}}^{-}\otimes1^{+}]_{\frac{3}{2}}\otimes1^{-}$
				&
				\\
				\hline
				$(0,1,0)$ &-0.137&0.058&0.015&0.080&0.354&0.088
				\\			
				\hline			
				&\multicolumn{3}{c|}{${^{2}P_{\frac{3}{2}}}$(${J^P={\frac{3}{2}}^{+}}$)}
				&\multicolumn{3}{c|}{${^{4}P_{\frac{3}{2}}}$(${J^P={\frac{3}{2}}^{+}}$)}
				&\multicolumn{1}{c}{${^{6}P_{\frac{3}{2}}}$(${J^P={\frac{3}{2}}^{+}}$)}  
				\\
				\cline{2-8}
				$(Y, I, I_3)$ 
				& ${\frac{1}{2}}^{-}\otimes0^{+}\otimes1^{-}$  
				& $[{\frac{1}{2}}^{-}\otimes1^{+}]_{\frac{1}{2}}\otimes1^{-}$      
				& $[{\frac{3}{2}}^{-}\otimes1^{+}]_{\frac{1}{2}}\otimes1^{-}$   
				& ${\frac{3}{2}}^{-}\otimes0^{+}\otimes1^{-}$    
				& $[{\frac{1}{2}}^{-}\otimes1^{+}]_{\frac{3}{2}}\otimes1^{-}$
				& $[{\frac{3}{2}}^{-}\otimes1^{+}]_{\frac{3}{2}}\otimes1^{-}$   
				& $[{\frac{3}{2}}^{-}\otimes1^{+}]_{\frac{5}{2}}\otimes1^{-}$
				\\
				\hline			
				$(0,1,0)$ &0.577&-0.030&0.098&0.152&0.508&0.157&0.242
				\\			
				\hline			
				&\multicolumn{3}{c|}{${^{4}P_{\frac{5}{2}}}$ (${J^P={\frac{5}{2}}^{+}}$)} &\multicolumn{1}{c|}{${^{6}P_{\frac{5}{2}}}$ (${J^P={\frac{5}{2}}^{+}}$)}  &\multicolumn{1}{c|}{${^{6}P_{\frac{7}{2}}}$ (${J^P={\frac{7}{2}}^{+}}$)}
				\\
				\cline{2-8}
				$ (Y, I, I_3)$  
				& ${\frac{1}{2}}^{-}\otimes1^{+}\otimes1^{-}$ 
				& ${\frac{3}{2}}^{-}\otimes0^{+}\otimes1^{-}$    
				& $[{\frac{3}{2}}^{-}\otimes1^{+}]_{\frac{3}{2}}\otimes1^{-}$  
				& $[{\frac{3}{2}}^{-}\otimes1^{+}]_{\frac{5}{2}}\otimes1^{-}$  
				& ${\frac{3}{2}}^{-}\otimes1^{+}\otimes1^{-}$
				\\
				\hline			
				$(0,1,0)$  &0.714&0.233&0.236&0.299&0.370
				\\
				\hline
				\bottomrule[1pt] 
				\multicolumn{8}{c}{$8_{2f}$: $\frac{1}{\sqrt2}\{ ({c}d)(\bar c[us])+({c}u)(\bar c[ds]) \}$}
				\\
				\hline
				&\multicolumn{2}{c|}{$^2S_{\frac{1}{2}}$($J^P={\frac{1}{2}}^{-}$)}
				&\multicolumn{1}{c|}{$^{4}S{{\frac{3}{2}}}$(${J^P={\frac{3}{2}}^{-}}$)}
				&\multicolumn{2}{c|}{${^{2}P_{\frac{1}{2}}}$(${J^P={\frac{1}{2}}^{+}}$)}
				&\multicolumn{1}{c|}{${^{4}P_{\frac{1}{2}}}$(${J^P={\frac{1}{2}}^{+}}$)}
				\\
				\cline{2-8} 
				$(Y, I, I_3)$  
				& ${\frac{1}{2}}^{-}\otimes0^{+}\otimes0^{+}$
				& ${\frac{1}{2}}^{-}\otimes1^{+}\otimes0^{+}$
				& ${\frac{1}{2}}^{-}\otimes1^{+}\otimes0^{+}$
				& ${\frac{1}{2}}^{-}\otimes0^{+}\otimes1^{-}$
				& $[{\frac{1}{2}}^{-}\otimes1^{+}]_{\frac{1}{2}}\otimes1^{-}$ 
				& $[{\frac{1}{2}}^{-}\otimes1^{+}]_{\frac{3}{2}}\otimes1^{-}$ 
				\\
				\hline			
				$(0,1,0)$ &-0.377&0.687&0.465&0.199&-0.184&0.235
				\\			
				\midrule[1pt]									
				&\multicolumn{2}{c|}{$^{2}P_{\frac{3}{2}}$ ($J^P={\frac{3}{2}}^{+}$)} &\multicolumn{1}{c|}{$^{4}P_{\frac{3}{2}}$ ($J^P={\frac{3}{2}}^{+}$)}   &\multicolumn{1}{c|}{${^{4}P_{\frac{5}{2}}^{+}}$ (${J^P={\frac{5}{2}}^{+}}$)}   
				\\
				\cline{2-8}
				$(Y, I, I_3)$ 
				& ${\frac{1}{2}}^{-}\otimes0^{+}\otimes1^{-}$
				& $[{\frac{1}{2}}^{-}\otimes1^{+}]_{\frac{3}{2}}\otimes1^{-}$ 
				& ${\frac{1}{2}}^{-}\otimes1^{+}\otimes1^{-}$
				& ${\frac{1}{2}}^{-}\otimes1^{+}\otimes1^{-}$ 
				\\
				\hline			
				$(0,1,0)$ &-0.307&0.803&0.381&0.558
				\\
				\hline				
				\bottomrule[1pt]
		\end{tabular}}
	\end{center}
\end{table*}
\begin{table*}[htbp]
	\caption{The magnetic moments of the pentaquark states in the diquark-triquark model with the wave function  $\frac{1}{\sqrt2}[({c}u)(\bar c\{ds\})-({c}d)(\bar c\{us\})]$ in $8_{1f}$ and $\frac{1}{\sqrt6}\{({c}d)(\bar c[us])-({ c}u)(\bar c[ds])-2( c s)(\bar c[ud])\}$ in $8_{2f}$ with isospin $(I,I_3) = (0,0)$.  The third line $J_{1}^{P_{1}}\otimes J_{2}^{P_{2}}\otimes J_{3}^{P_{3}}$ are corresponding to the angular momentum and parity of triquark, diquark and orbital, respectively. The unit is the magnetic moments of the proton.}\label{lll}
	\begin{center}
		\resizebox{0.90\columnwidth}{!}{
			\begin{tabular}{c|c|c|c|c|c|c|c} \toprule[1pt]	
				\hline
				\multicolumn{8}{c}{$8_{1f}$: $\frac{1}{\sqrt2}[({c}u)(\bar c\{ds\})-({c}d)(\bar c\{us\})]$}
				\\
				\hline
				&\multicolumn{3}{c|}{$^{2}S_{\frac{1}{2}}$($J^P={\frac{1}{2}}^{-}$)}
				&\multicolumn{3}{c|}{${^{4}S_{\frac{3}{2}}}$(${J^P={\frac{3}{2}}^{-}}$)}
				&\multicolumn{1}{c}{$^{6}S_{\frac{5}{2}}^{-}$($J^P={\frac{5}{2}}^{-}$)} 
				\\
				\cline{2-8} 
				$(Y, I, I_3)$ 
				& ${\frac{1}{2}}^{-}\otimes0^{+}\otimes0^{+}$
				& ${\frac{1}{2}}^{-}\otimes1^{+}\otimes0^{+}$  
				& ${\frac{3}{2}}^{-}\otimes1^{+}\otimes0^{+}$      
				& ${\frac{1}{2}}^{-}\otimes1^{+}\otimes0^{+}$ 
				& ${\frac{3}{2}}^{-}\otimes0^{+}\otimes0^{+}$
				& ${\frac{3}{2}}^{-}\otimes1^{+}\otimes0^{+}$  
				& ${\frac{3}{2}}^{-}\otimes1^{+}\otimes0^{+}$
				\\
				\hline				
				$(0,0,0)$ &0.033&0.574&-0.601&0.910&-0.555&-0.056&0.352
				\\
				\hline			
				&\multicolumn{3}{c|}{${^{2}P_{\frac{1}{2}}}$ (${J^P={\frac{1}{2}}^{+}}$)} &\multicolumn{3}{c|}{${^{4}P_{\frac{1}{2}}}$ (${J^P={\frac{1}{2}}^{+}}$)}
				&  
				\\
				\cline{2-8}
				$(Y, I, I_3)$ 
				& ${\frac{1}{2}}^{-}\otimes0^{+}\otimes1^{-}$  
				& $[{\frac{1}{2}}^{-}\otimes1^{+}]_{\frac{1}{2}}\otimes1^{-}$      
				& $[{\frac{3}{2}}^{-}\otimes1^{+}]_{\frac{1}{2}}\otimes1^{-}$  
				& ${\frac{3}{2}}^{-}\otimes0^{+}\otimes1^{-}$    
				& $[{\frac{1}{2}}^{-}\otimes1^{+}]_{\frac{3}{2}}\otimes1^{-}$  
				& $[{\frac{3}{2}}^{-}\otimes1^{+}]_{\frac{3}{2}}\otimes1^{-}$
				&
				\\
				\hline
				$(0,0,0)$ &0.062&-0.126&0.265&0.473&-0.345&-0.064
				\\				
				\hline			
				&\multicolumn{3}{c|}{${^{2}P_{\frac{3}{2}}}$(${J^P={\frac{3}{2}}^{+}}$)}
				&\multicolumn{3}{c|}{${^{4}P_{\frac{3}{2}}}$(${J^P={\frac{3}{2}}^{+}}$)}
				&\multicolumn{1}{c}{${^{6}P_{\frac{3}{2}}}$	(${J^P={\frac{3}{2}}^{+}}$)}  \\
				\cline{2-8}
				$(Y, I, I_3)$ 
				& ${\frac{1}{2}}^{-}\otimes0^{+}\otimes1^{-}$  
				& $[{\frac{1}{2}}^{-}\otimes1^{+}]_{\frac{1}{2}}\otimes1^{-}$      
				& $[{\frac{3}{2}}^{-}\otimes1^{+}]_{\frac{1}{2}}\otimes1^{-}$   
				& ${\frac{3}{2}}^{-}\otimes0^{+}\otimes1^{-}$    
				& $[{\frac{1}{2}}^{-}\otimes1^{+}]_{\frac{3}{2}}\otimes1^{-}$
				& $[{\frac{3}{2}}^{-}\otimes1^{+}]_{\frac{3}{2}}\otimes1^{-}$   
				& $[{\frac{3}{2}}^{-}\otimes1^{+}]_{\frac{5}{2}}\otimes1^{-}$
				\\
				\hline				
				$(0,0,0)$ &0.143&0.671&-0.503&0.707&-0.363&-0.002&0.212
				\\			
				\hline
				&\multicolumn{3}{c|}{${^{4}P_{\frac{5}{2}}}$ (${J^P={\frac{5}{2}}^{+}}$)} &\multicolumn{1}{c|}{${^{6}P_{\frac{5}{2}}}$ (${J^P={\frac{5}{2}}^{+}}$)}  &\multicolumn{1}{c|}{${^{6}P_{\frac{7}{2}}}$ (${J^P={\frac{7}{2}}^{+}}$)}
				\\
				\cline{2-8}
				$ (Y, I, I_3)$   
				& ${\frac{1}{2}}^{-}\otimes1^{+}\otimes1^{-}$ 
				& ${\frac{3}{2}}^{-}\otimes0^{+}\otimes1^{-}$    
				& $[{\frac{3}{2}}^{-}\otimes1^{+}]_{\frac{3}{2}}\otimes1^{-}$  
				& $[{\frac{3}{2}}^{-}\otimes1^{+}]_{\frac{5}{2}}\otimes1^{-}$  
				& ${\frac{3}{2}}^{-}\otimes1^{+}\otimes1^{-}$
				\\
				\hline
				
				$(0,0,0)$  &1.008&-0.445&0.041&0.313&0.420
				\\
				\hline
				\bottomrule[1pt]
				\multicolumn{8}{c}{$8_{2f}$: $\frac{1}{\sqrt6}\{({c}d)(\bar c[us])-({ c}u)(\bar c[ds])-2( c s)(\bar c[ud])\}$}
				\\
				\hline
				&\multicolumn{2}{c|}{$^2S_{\frac{1}{2}}$($J^P={\frac{1}{2}}^{-}$)}
				&\multicolumn{1}{c|}{$^{4}S{{\frac{3}{2}}}$(${J^P={\frac{3}{2}}^{-}}$)}
				&\multicolumn{2}{c|}{${^{2}P_{\frac{1}{2}}}$(${J^P={\frac{1}{2}}^{+}}$)}
				&\multicolumn{1}{c|}{${^{4}P_{\frac{1}{2}}}$(${J^P={\frac{1}{2}}^{+}}$)}
				\\
				\cline{2-8} 
				$(Y, I, I_3)$  
				& ${\frac{1}{2}}^{-}\otimes0^{+}\otimes0^{+}$
				& ${\frac{1}{2}}^{-}\otimes1^{+}\otimes0^{+}$
				& ${\frac{1}{2}}^{-}\otimes1^{+}\otimes0^{+}$
				& ${\frac{1}{2}}^{-}\otimes0^{+}\otimes1^{-}$
				& $[{\frac{1}{2}}^{-}\otimes1^{+}]_{\frac{1}{2}}\otimes1^{-}$ 
				& $[{\frac{1}{2}}^{-}\otimes1^{+}]_{\frac{3}{2}}\otimes1^{-}$ 
				\\
				\hline			
				$(0,0,0)$ &-0.377&0.223&-0.231&0.164&-0.035&-0.165
				\\			
				\hline									
				&\multicolumn{2}{c|}{$^{2}P_{\frac{3}{2}}$ ($J^P={\frac{3}{2}}^{+}$)} &\multicolumn{1}{c|}{$^{4}P_{\frac{3}{2}}$ ($J^P={\frac{3}{2}}^{+}$)}   &\multicolumn{1}{c|}{${^{4}P_{\frac{5}{2}}^{+}}$(${J^P={\frac{5}{2}}^{+}}$}\\
				\cline{2-8}
				$(Y, I, I_3)$ 
				& ${\frac{1}{2}}^{-}\otimes0^{+}\otimes1^{-}$
				& $[{\frac{1}{2}}^{-}\otimes1^{+}]_{\frac{3}{2}}\otimes1^{-}$ 
				& ${\frac{1}{2}}^{-}\otimes1^{+}\otimes1^{-}$
				& ${\frac{1}{2}}^{-}\otimes1^{+}\otimes1^{-}$ 
				\\
				\hline			
				$(0,0,0)$ &-0.359&0.293&-0.165&-0.196
				\\	
				\hline				
				\bottomrule[1pt]
		\end{tabular}}
	\end{center}
\end{table*}

I have compared the magnetic moment of $P_{cs}(4459)$ in three configurations as shown in the following Table \ref{gao}. The magnetic moment and numerical results illustrate that molecular model is distinguishable from the other two models in $0({\frac{1}{2}}^{-})$ but it is indistinguishable in $0({\frac{3}{2}}^{-})$. As far as diquark-diquark-antiquark model and diquark-triquark model are concerned, they are completely indistinguishable in $0({\frac{1}{2}}^{-})$ and $0({\frac{3}{2}}^{-})$.  
In addition to this, the magnetic moment of $P_{cs}(4459)$ have been studied in other papers. In Ref.~\cite{Li:2021ryu}, the numerical value in the molecular picture was obtained as $\mu_{P_{cs}} = -0.062\mu_{N}$ with 0(${\frac{1}{2}}^{-}$) and $\mu_{P_{cs}} = 0.465\mu_{N}$ with 0(${\frac{3}{2}}^{-}$). In Ref.\cite{Ozdem:2021ugay}, the magnetic dipole moment of $P_{cs}(4459)$ in the molecular and diquark-diquark-antiquark pictures are extracted as $\mu_{P_{cs}} = 1.75\mu_{N}$ and $\mu_{P_{cs}}=0.34\mu_{N}$. These numerical results differ from our results $\mu_{P_{cs}} = -0.531\mu_{N}$ with 0(${\frac{1}{2}}^{-}$) and $\mu_{P_{cs}} = -0.231\mu_{N}$ with 0(${\frac{3}{2}}^{-}$) in the molecular model and $\mu_{P_{cs}}=0.223\mu_{N}$ in diquark-diquark-antiquark model because of the wavefunction and the quark mass. We compare the results in Table \ref{g11}

\begin{table}
	\centering
	\caption{The magnetic moments of the $P_{cs}(4459)$ in the molecular model, the diquark-diquark-antiquark model and the diquark-triquark model in $8_{2f}$ representation with isospin $(I,I_3) = (0,0)$.}\label{gao}
	\begin{center}
	\resizebox{\columnwidth}{!}{
		\begin{tabular}{|c|c|c|c|c|c|}
			\toprule[1pt]	
			\hline
			$P_{cs}(4459)$ & Multiplet & Spin-orbit coupling& $I(J^P)$ & Magnetic moment & Numerical results
			\\
			\hline
			\multirow{2}{*}{Molecular model}
			& \multirow{2}{*}{$8_{2f}$}
			& \multirow{2}{*}{${\frac{1}{2}}^{+}\otimes1^{-}\otimes0^{+}$}
			& $0({\frac{1}{2}}^{-})$
			& $\frac{1}{9}(6\mu_{\bar{c}}-3\mu_{c}+\mu_{u}+\mu_{d}+4\mu_{s})$	
			& -0.531
			\\
			\cline{4-6} 	
			& &
			& $0({\frac{3}{2}}^{-})$
			& $\frac{1}{6}(6\mu_{c}+6\mu_{\bar{c}}+\mu_{u}+\mu_{d}+4\mu_{s})$
			& -0.231
			\\
			\hline
			\multirow{2}{*}{diquark-diquark-antiquark model}
			& \multirow{2}{*}{$8_{2f}$}
			& \multirow{2}{*}{$1^{+}\otimes0^{+}\otimes{\frac{1}{2}}^{-}\otimes0^{+}$}
			& $0({\frac{1}{2}}^{-})$
			& $\frac{1}{9}(6\mu_{c}-3\mu_{\bar{c}}+\mu_{u}+\mu_{d}+4\mu_{s})$	
			& 0.223
			\\
			\cline{4-6} 	
			& &
			& $0({\frac{3}{2}}^{-})$
			& $\frac{1}{6}(6\mu_{c}+6\mu_{\bar{c}}+\mu_{u}+\mu_{d}+4\mu_{s})$
			& -0.231
			\\
			\hline	\multirow{2}{*}{diquark-triquark model}
			& \multirow{2}{*}{$8_{2f}$}
			& \multirow{2}{*}{${\frac{1}{2}}^{-}\otimes1^{+}\otimes0^{+}$}
			& $0({\frac{1}{2}}^{-})$
			& $\frac{1}{9}(6\mu_{c}-3\mu_{\bar{c}}+\mu_{u}+\mu_{d}+4\mu_{s})$	
			& 0.223
			\\
			\cline{4-6} 	
			& &
			& $0({\frac{3}{2}}^{-})$
			& $\frac{1}{6}(6\mu_{c}+6\mu_{\bar{c}}+\mu_{u}+\mu_{d}+4\mu_{s})$
			& -0.231
			\\
			\hline
			\bottomrule[1pt]
	\end{tabular}}
 \end{center}
\end{table}	
\begin{table}[htbp]
	\centering
	\caption{Our results and other theoretical results for the magnetic moment of $P_{cs}(4459)$.The unit is the magnetic moments of the proton. The  A, B, and C are corresponding to the molecular model, diquark-diquark-antiquark model and  diquark-triquark model}\label{g11}
	\begin{center}
		\resizebox{0.6\columnwidth}{!}{
			\begin{tabular}{c|c|c|c|c|c|c}
				\hline
				Cases & \multicolumn{2}{c|}A &\multicolumn{2}{c|}{B} &\multicolumn{2}{c}{C} 
				\\
				\hline
				$J^P$&${\frac{1}{2}}^{-}$&${\frac{3}{2}}^{-}$&${\frac{1}{2}}^{-}$&${\frac{3}{2}}^{-}$&${\frac{1}{2}}^{-}$&${\frac{3}{2}}^{-}$
				\\
				\hline
				Our results & -0.531&-0.231	& 0.223 &-0.231&0.223&-0.231
				\\
				\hline
				Ref.\cite{Li:2021ryu} & -0.062&0.465	& - &-&-&-
				\\
				\hline
				Ref.\cite{Ozdem:2021ugay} & 1.75&-	& 0.34 &-&-&-
				\\
				\hline
		\end{tabular}}
	\end{center}
\end{table}	

	\section{SUMMARY}
	\label{sec5}
	Inspired by the recently observed $P_{cs}(4459)$, we systematically calculates the magnetic moments of $P_{cs}$ with $J^{P}={\frac{1}{2}}^{\pm}, {\frac{3}{2}}^{\pm}, {\frac{5}{2}}^{\pm}$, and ${\frac{7}{2}}^{+}$ in three models: molecular, diquark-diquark-antiquark, and diquark-triquark.  Comparing the numerical results of the above three models, we observe that the magnetic moments of the states with the same quantum numbers are different. Indeed, even within the same model, the magnetic moments with different configurations are different. Next, we compare the magnetic moment of $P_{cs}(4459)$ in three configurations, which has been predicted involves an S-wave state with $I(J^P)=0({\frac{1}{2}}^{-})$ and  $I(J^P)=0({\frac{3}{2}}^{-})$. The result shows that the molecular model is different from the other two models in  $I(J^P)=0({\frac{1}{2}}^{-})$. These findings highlight that magnetic moments are helpful to determine their internal structures when the experimental data of $P_{cs}$ keeps accumulating, since the magnetic moments encode information about the charge distributions.

	\section*{Acknowledgments}
	This project is supported by the National Natural Science Foundation of China under
	Grants No. 11905171 and No. 12047502. This work is also supported by the Natural Science Basic Research Plan in Shaanxi Province of China (Grant No. 2022JQ-025).

\end{document}